\documentclass[onecolumn,fleqn,usenatbib]{mnras}

\usepackage{amsmath}	
\usepackage{amssymb}	
\usepackage{txfonts} 

\usepackage[T1]{fontenc}

\DeclareRobustCommand{\VAN}[3]{#2}
\let\VANthebibliography\thebibliography
\def\thebibliography{\DeclareRobustCommand{\VAN}[3]{##3}\VANthebibliography}

\usepackage{ulem}

\usepackage[dvipdfmx]{graphicx}	

\title[Magnetic field dependence of neutrino-driven SNe]{Two-dimensional numerical study for magnetic field dependence of neutrino-driven core-collapse supernova models}

\author[J. Matsumoto et al.]{J. Matsumoto$^{1}$\thanks{Email:jin.matsumoto@fukuoka-u.ac.jp, jin@kusastro.kyoto-u.ac.jp},
T. Takiwaki$^{2}$, K. Kotake$^{1, 3}$, Y. Asahina$^{4}$ and H. R. Takahashi$^{5}$\\
$^{1}$Research Institute of Stellar Explosive Phenomena, Fukuoka University, Fukuoka 814-0180, Japan\\
$^{2}$Division of Science, National Astronomical Observatory of Japan, Tokyo 181-8588, Japan\\
$^{3}$Department of Applied Physics, Fukuoka University, Fukuoka 814-0180, Japan\\
$^{4}$Center for Computational Sciences, Tsukuba University, Ibaraki 305-8577, Japan\\
$^{5}$Faculty of Arts and Sciences, Department of Natural Sciences, Komazawa University, Tokyo 154-8525, Japan}

\date{Accepted 2020 October 06. Received 2020 October 05; in original form 2020 August 20}

\pubyear{2020}

\begin{document}
\label{firstpage}
\pagerange{\pageref{firstpage}--\pageref{lastpage}}
\maketitle

\begin{abstract}
We study the effects of the magnetic field on the dynamics of non-rotating
stellar cores  by performing two-dimensional (2D), magnetohydrodynamics (MHD)
simulations. To this end, we have updated our neutrino-radiation-hydrodynamics
supernova code to include MHD employing a divergence cleaning method with both
careful treatments of finite volume and area reconstructions. By changing the
initial strength of the magnetic field, the evolution of $15.0$, $18.4$ and
$27.0$ $M_\odot$ presupernova progenitors is investigated. An intriguing
finding in our study is that the neutrino-driven explosion occurs regardless
of the strength of the initial magnetic field. For the 2D models  presented
in this work, the neutrino heating is the main driver for the explosion,
whereas the magnetic field secondary contributes to the pre-explosion dynamics.
Our results show that the strong magnetic field weakens the growth of the
neutrino-driven turbulence in the small scale compared to the weak magnetic
field. This results in the slower increase of  the turbulent kinetic energy
in the postshock region, leading to the slightly delayed onset of the
shock revival for models with the stronger initial magnetic field.
\end{abstract}



\section{Introduction}
Core-collapse supernovae (CCSNe) are one of the most energetic explosions
in the universe, marking the catastrophic end of massive stars. Extensive
investigations over the decades (e.g. \citealt{Colgate66}, see also
\citealt{Liebendorfer05, Tony12, Janka12, Kotake12, Burrows13, Foglizzo15, Mueller20}
for reviews) have shown that the most promising way to explode massive stars
($\gtrsim 8 M_{\odot}$) is the neutrino mechanism \citep{bethe}. In this
mechanism, neutrinos from the protoneutron star (PNS) heat the material
behind the stalled shock, leading to the shock revival into explosion.
The neutrino heating efficiency is significantly enhanced by non-radial
flows triggered by various hydrodynamic instabilities including
neutrino-driven/PNS convection and the standing accretion shock instability
(SASI; \citealt{Blondin03,Foglizzo06}). This has been confirmed in a growing
number of self-consistent CCSN simulations (e.g.  \citealt{Hanke13,Takiwaki14,lentz15,Bernhard15,Summa16,Takiwaki16,O'Connor18,Pan18,Ott18,Kuroda18,Nagakura19,Vartanyan19,Nakamura19,Melson20}),
some of which could closely account for canonical CCSNe with the explosion
energies of the order of $10^{51}$ erg ($\equiv$ 1 Bethe, 1 B in short)
or less.

However, there are some very energetic subclasses of supernovae, which are
highly unlikely to be explained by the conventional neutrino mechanism.
Those events so far observed include hypernovae \citep[e.g.][]{Iwamoto98,Soderberg06, Nomoto10}
and superluminous SNe \citep[SLSNe; see, e.g.][for reviews]{Gal-Yam12,Nicholl13,Moriya18}.
The most plausible scenario to account for these extreme events requires
additional energy injection via the magnetohydrodynamically-driven (MHD, in short)
explosions (e.g. \citealt{Wheeler02,Burrows07,Dessart08,Dessart12}).
The kinetic energy of hypernovae exceeds $\sim 10$ B, which is ten times
larger than that of the canonical CCSNe\footnote{In order to explain
observational features of HNe, an extensive study has been carried out
so far in various contexts (e.g. \citealt{Maeda03,Woosley06,Tominaga09,Mazzali14,greiner15,Metzger15}
for references therein).}. Concerning the SLSNe, the luminosity is $10$--$100$
times higher than that of the typical CCSNe. To explain the excess of
the bright luminosity, several scenarios have been proposed including
the interaction scenario between the SN ejecta and its dense circumstellar
medium \citep{Chevalier11, Moriya13, Sorokina16} or the pair-instability
SN scenario \citep{Barkat67, Rakavy67, Terreran17}. Besides, the injection
of the additional energy into the SN ejecta by the central engine of
a rapidly rotating proto-magnetar is also a promising scenario for the
excess of luminosity \citep{Kasen10, Woosley10,Wang15,Chen16}. For the two
classes of the extreme events mentioned above, one could speculate that
a common solution requires the strong (magnetar-class), large-scale magnetic
fields, which  can directly couple the newly-born PNS (or proto-magnetar) to
its surroundings via various MHD processes \citep{usov,thompson94,bucci09,metzger11}.
 
The MHD explosion mechanism originally proposed in the 1970s
\citep{Bisnovatyi-Kogan70,LeBlancWilson70,Meier76, EMuller79} has recently
received considerable attention. To study the MHD mechanism in various
contexts, extensive numerical simulations have been conducted so far
(e.g. \citealt{Symbalisty84, Ard00, Kotake04, Sawai05, Shibata06, Sergey06,
Suwa07, Takiwaki09, Takiwaki11, Obergaulinger06a, Obergaulinger06b,
Winteler12,Sawai14,Sawai16, Obergaulinger17,  Obergaulinger20, Obergaulinger18,
Moesta14,Moesta15,Bugli20,Kuroda20}). However, the extremely huge explosion
energy ($\sim 10$ B) has yet to be obtained in these simulations. And there
still remains a big issue whether a combination of strong magnetic field
and rapid rotation can be achieved in the precollapse iron core.
 
Assuming the magnetic flux conservation, the strong surface magnetic field
($\sim 1$ kG) of OB-type stars (e.g. \citealt{Donati02, Donati06, Hubrig06})
is considered as a possible candidate to account for the magnetic field of
magnetars. On the other hand, the majority of massive stars possess the
weak magnetic field, which is supported by both the observation
\citep[$\lesssim 100$ G;][]{Wade15} and stellar evolution calculations
\citep{Heger05}. In the latter case, enough amplification of the magnetic
field during and after the collapse of the massive star is necessary to
facilitate the MHD mechanism. Since the magnetic field amplification due to
the field wrapping as the consequence of rapid and  strong differential
rotation scales linearly in time and takes a long time compared to the
dynamical time scale of the system, the drastic and efficient field
amplification mechanism is required when the seed field of the massive
star is adequately weak. The exponential growth of the magnetic field is
beneficial to amplify the weak seed field of the stellar core to the
dynamically relevant level. The magnetic field amplification due to the
magnetorotational instability \citep[MRI;][]{Balbus91} is a candidate for
the efficient field amplification mechanism in the rotating stellar cores
\citep{Akiyama03, Masada06, Masada07, Masada12, Masada15, Obergaulinger09,
Sawai13, Sawai14, Sawai16, Guilet15a, Guilet15b, Moesta15, Rembiasz16, Reboul20}.

On the other hand, stellar evolution calculations pointed out that the
majority of the magnetic core of the massive star is expected to be rotating
slowly at the pre-collapse stage \citep{Heger05, Ott06, Langer12}. Especially,
the stronger magnetic field would lead to the more efficient angular momentum
loss of the stellar core by the magnetic braking even if some of the stars
rapidly rotate initially \citep{Ramrez-Agudelo13} or the stars experience
spin-up due to stellar mergers \citep{Chatzopoulos20}. Asteroseismology of
low-mass stars also suggested that an unmodeled, more efficient angular
momentum transport process is necessary to explain the spin period of the
cores \citep{Cantiello14,Fuller14}. Observations of surface rotational
velocities of B-type stars with strong magnetic fields also favor slow
rotators \citep{Shultz18}. Even in such slowly rotating progenitors, it has
been pointed out that the precollapse magnetic field, if sufficiently strong,
could affect the explosion dynamics. In the context of the slowly- and
non-rotating progenitor, \citet{Endeve10, Endeve12} and \citet{Obergaulinger14}
studied the dynamics of the MHD core-collapse and the exponential amplification
of the magnetic field due to the SASI and convection. More recently,
\citet{Muller20b} have addressed the role of the magnetic field in the
neutrino-driven explosion by performing three-dimensional (3D) MHD
simulations of a slowly rotating progenitor of a $15 M_{\odot}$ star.

Joining in these efforts to study the impact of the magnetic field on both
the extreme and ordinary explosions of the massive stars, we investigate
the magnetic field dependence of neutrino-driven explosion in the
{\it non-rotating} cores by performing two-dimensional (2D), axisymmetric,
MHD core-collapse simulations for several representative progenitors.
In the context of 2D simulations of the non-rotating and magnetized cores,
\cite{Obergaulinger14} were the first to point out that magnetic pressure
support in the gain region (via turbulence) fosters the onset of
neutrino-driven explosion. This result clearly presented evidence that
implementation of appropriate neutrino transport is needed for a quantitative
study of MHD CCSN modeling. However, only electron and anti-electron neutrinos
were taken into account in \citet{Obergaulinger14} at that time (see, however,
\citealt{Obergaulinger20}). To revisit the problem, we have updated our
supernova code (3DnSNe) to include MHD by implementing a divergence cleaning
method \citep{Dedner02} with both finite volume and area reconstructions
based on \citet{Mignone14}. Our base code deals with three-flavor neutrino
transport (namely, $\nu_e,\bar{\nu}_e,\nu_X$ with $\nu_X$ denoting the
heavy-lepton neutrinos) \citep{Kotake18} based on the Isotropic Diffusion
Source Approximation (IDSA) scheme \citep{Liebendorfer09}, in which a detailed
code comparison was already made in spherically symmetric simulations
\citep{OConnor18}, in 2D simulations \citep{Kotake18} using a widely
used 20 $M_{\odot}$ star of \citet{Woosley07} and in 3D simulations 
\citep{Cabezon18} using 15 $M_{\odot}$ stars of \citet{Woosley95} and \citet{Woosley07}.

This paper is organized as follows: In Section~\ref{numerical methods},
the numerical methods and models are described. Our numerical results
of the MHD core-collapse of non-rotating stellar cores in axisymmetry
are presented in Section~\ref{results}. In Section~\ref{B-field of PNS},
we discuss the field configuration of the proto-magnetar based on our
simulation results. Finally, we summarize and discuss our findings in
Section~\ref{summary}.

\section{Numerical methods and models} \label{numerical methods}
We have updated our supernova code, 3DnSNe\footnote{Previously, \cite{Takiwaki12}
and \cite{Takiwaki14} employed the {\small ZEUS-MP} code \citep{Hayes06}
where the (tensor-type) artificial viscosity was used to capture the shock
(see also \citealt{Iwakami08}). \cite{Suwa10} and \cite{Suwa16} utilized
the {\small ZEUS-2D} code of \citet{Stone92}.} \citep{Takiwaki16}, that is
designed for CCSN simulations in a 3D spherical coordinate system to the
latest version. In this work, the code is now extended to an MHD code from
a hydrodynamic (HD) one with spectral neutrino transport that is solved
by the IDSA scheme \citep{Liebendorfer09}. We have updated the original
(two-flavor, i.e. $\nu_e,\bar{\nu}_e$) IDSA scheme in several manners,
such that the evolution of the streaming neutrinos is self-consistently
solved \citep{Takiwaki14} and that three-flavor neutrino transport is solved
including approximate general relativistic corrections (e.g. \citealt{Kotake18}
for more details). A detailed code comparison has been performed in
\citet{OConnor18} with one-dimensional (1D) geometry. The 3DnSNe code
has been used in the following works: \citet{Cherry20, Zaizen20, Sasaki20,
Nakamura19, Sasaki17, Sotani16, Sotani20, Nakamura15}.

In the latest code, we solve the ideal MHD equations in the spherical
coordinate system ($r$, $\theta$, $\phi$). The governing equations are
\begin{eqnarray}
\frac{\partial \rho}{\partial t} + \nabla \cdot (\rho \mbox{\boldmath $v$}) = 0 \;, \label{basic eq1}
\end{eqnarray}
\begin{eqnarray}
\frac{\partial (\rho \mbox{\boldmath $v$})}{\partial t} 
+ \nabla \cdot (\rho \mbox{\boldmath $v$} \mbox{\boldmath $v$} + P_t {\bf I}
- \mbox{\boldmath $B$} \mbox{\boldmath $B$}) = - \rho \nabla \Phi \;, \label{basic eq2}
\end{eqnarray}
\begin{eqnarray}
\frac{\partial e}{\partial t} 
+ \nabla \cdot [(e + P_t) \mbox{\boldmath $v$} - 
\mbox{\boldmath $B$} (\mbox{\boldmath $v$} \cdot \mbox{\boldmath $B$})]
= - \rho \mbox{\boldmath $v$} \cdot \nabla \Phi + Q_E \;, \label{basic eq3}
\end{eqnarray}
\begin{eqnarray}
\frac{\partial \mbox{\boldmath $B$}}{\partial t} 
+ \nabla \cdot (\mbox{\boldmath $v$} \mbox{\boldmath $B$} 
- \mbox{\boldmath $B$} \mbox{\boldmath $v$} + \psi {\bf I}) = 0 \;, \label{basic eq4}
\end{eqnarray}
\begin{eqnarray}
\frac{\partial \psi}{\partial t} + c_h^2 \nabla \cdot \mbox{\boldmath $B$}
= - \frac{c_h^2}{c_p^2} \psi \;, \label{basic eq5}
\end{eqnarray}
\begin{eqnarray}
\frac{\partial \rho Y_l}{\partial t} + \nabla \cdot (\rho Y_l  \mbox{\boldmath $v$}) = \Gamma_l \;, \label{basic eq6}
\end{eqnarray}
\begin{eqnarray}
\frac{\partial \rho Z_m}{\partial t} + \nabla \cdot (\rho Z_m  \mbox{\boldmath $v$}) + \frac{\rho Z_m}{3}\nabla \cdot \mbox{\boldmath $v$}  = Q_m \;, \label{basic eq7}
\end{eqnarray}
\begin{eqnarray}
\Delta \Phi = 4 \pi G \rho \;, \label{basic eq8}
\end{eqnarray}
where $\rho$, $\mbox{\boldmath $v$}$, $\mbox{\boldmath $B$}$, $P_t$, $e$
and $\Phi$ are the mass density, the fluid velocity vector, the magnetic
field vector, the total (thermal and magnetic) pressure, the total energy
density and the gravitational potential, respectively. $Y_l$ is the lepton
fraction and the subscription $l$ denotes the species of leptons: $l=e,
\nu_e, \bar{\nu}_e, \nu_X$ and $Z_m$ is the specific internal energy of
the trapped neutrinos and $m$ represents the species of neutrinos:
$m=\nu_e,\bar{\nu}_e,\nu_X$. $Q_E$, $Q_{m}$ are the change of the energy
and $\Gamma_l$ is the change of number fraction due to the interaction
with the fluid and neutrinos. ${\bf I}$ is the unit matrix. The explicit
expressions of the governing equations in the spherical polar coordinates
are given in Appendix \ref{equations in spherical coordinate}. 

As for the approximate Riemann solver, the HLLD scheme \citep{Miyoshi05}
is newly implemented in our code to solve equations~(\ref{basic eq1})--(\ref{basic eq4})
in a conservative form (see Appendix~\ref{equations in spherical coordinate}
for the treatment of equations \ref{basic eq6} and \ref{basic eq7}).
To get around the Carbuncle instability, we switch from the HLLD scheme
to the HLLE scheme \citep{einfeldt} in the vicinity of strong shocks
(see \citealt{Kim03} and the references therein). Equation (\ref{basic eq5})
and $\psi$ in the induction equation (\ref{basic eq4}) are related to
the divergence cleaning method proposed by \citep{Dedner02}. This method
reduces numerical errors of the solenoidal property of the magnetic
field within minimal levels. The average relative divergence error estimated
by Error$3$($\mbox{\boldmath $B$}$)$^{ave}$ proposed by \citet{Zhang16}
is less than 1\% in our work. $c_h$ and $c_p$ characterize the propagation
speed and the damping rate of the numerical divergence of the magnetic field,
respectively (see Appendix~\ref{discretization} for details). Equation
\eqref{basic eq5} is solved using HLLE scheme. To retain the total energy
including gravitational binding energy, we use the method of \cite{Muller10}
in solving equation \eqref{basic eq3}. In equation \eqref{basic eq8},
the spherically symmetric gravitational potential is taken in the form of
phenomenological general relativistic potential of Case A in \cite{Marek06} 
and the multi-pole components are added following \citet{Wongwathanarat2010}.
 
We use a finite volume method \citep{Li03, Mignone14} to solve conservation
equations basically. However, since the divergence cleaning method is
employed in our code, both the finite volume and area methods are used
to solve the induction equation. The details of this special treatment,
in addition to the reconstructions of the physical variables for the
second-order accuracy in space, are described in  Appendix~\ref{discretization}
and~\ref{reconstruction}. A code test is given in Appendix~\ref{code test1}.

Our setups for microphysics are similar to those of \cite{OConnor18}.
The adopted neutrino reaction rate is set5a of \cite{Kotake18}, i.e.
the weak magnetism and recoil correction \citep{Horowitz02} as well as 
nucleon-nucleon bremsstrahlung is added to the standard opacity set
of \cite{Bruenn85}. In this run, 20 energy groups that logarithmically
spread from 1 to 300 MeV are employed. We use the equation of state (EOS)
by \cite{Lattimer91} (incompressibility K = 220 MeV).

We employ the non-rotating presupernova progenitors of $15.0$, $18.4$ and
$27.0$ $M_\odot$ of \citet{Woosley02}. As for the initial configuration of
the magnetic fields, we assume a simple topology following \citet{Suwa07,
Takiwaki14, Obergaulinger14}. The magnetic field is given by a vector
potential in the $\phi$-direction of the form
\begin{eqnarray}
A_{\phi} = \frac{B_0}{2} \frac{r^3_0}{r^3 + r^3_0} r \sin \theta \;, \label{eq: initial magnetic field}
\end{eqnarray}
where $r_0 = 1000$ km characterizes the topology of the field.
The magnetic field is uniform when the radius, $r$, is smaller than $r_0$,
while it is like dipole field when $r$ is larger than $r_0$. $B_0$ determines
the strength of the magnetic field inside the core ($r < r_0$). In this study,
we set $B_0 = 10^{10}$, $10^{11}$ or $10^{12}$ G. The model name is labelled as
`s27.0B10', which represents the $27.0$ $M_\odot$ model with $B_0=10^{10}$ G.
We choose s27.0B10 as a fiducial model because 2D (albeit, non-magnetized)
results using this progenitor are available in the literature (e.g.
\citealt{Hanke13,Summa16}). We follow the dynamics up to $t_{\rm fin}
\sim 400 - 500$ ms after bounce, depending on the progenitor models.
In most of the models, we terminate the simulations at the final time
seeing that the diagnostic explosion energies are greater than $10^{50}$ erg.
We leave the more long-term simulation for future work.

The calculations are performed in axisymmetry. Therefore, the derivatives
with respect to the $\phi$-direction (i.e. $\frac{\partial}{\partial \phi}$)
are taken to be zero in the governing equations when we run 2D simulations.
The grid spacing in this work is the similar to that of 2D runs in
\citet{Takiwaki14}. In the radial direction, a logarithmically stretched
grid is adopted for $480$ zones that cover from the center up to $5000$ km,
whereas the polar angle in the $\theta$-direction is uniformly divided into
$\Delta \theta = \pi / 128$. The innermost $10$ km are computed in spherical
symmetry to avoid excessive time-step limitations. Reflective boundary
conditions are imposed on the inner radial boundary ($r=0$), while fixed-boundary
conditions are adopted for the outer radial boundary ($r=5000$ km) except
the gravitational potential that is inversely proportional to the radius
at outer ghost cells. A reflecting boundary condition is imposed on the 2D
symmetry axis (e.g. the $z$-axis in our 2D run). A numerical resolution test
is given in Appendix \ref{resolution study}.

\begin{figure}
\begin{center}
\scalebox{0.9}{{\includegraphics{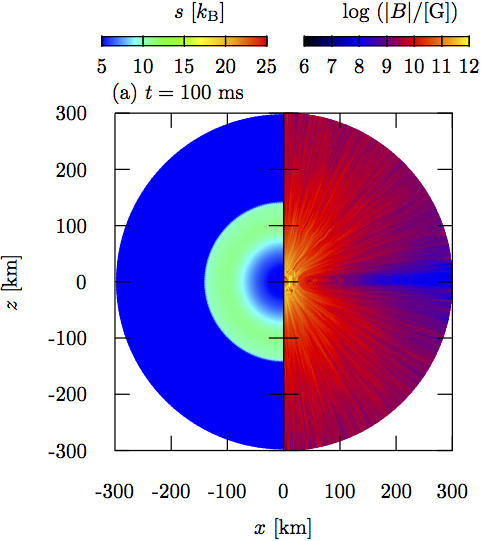}}}
\scalebox{0.9}{{\includegraphics{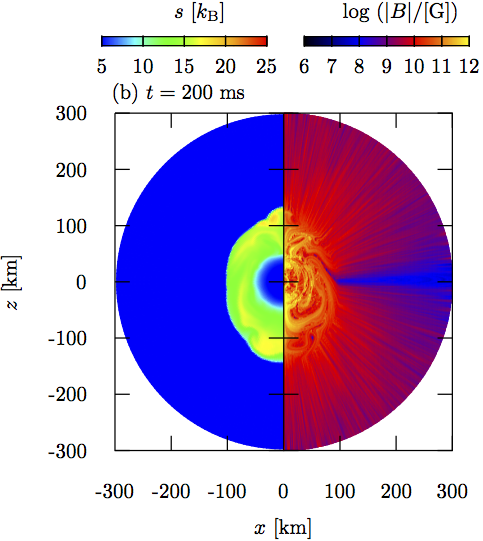}}}
\scalebox{0.9}{{\includegraphics{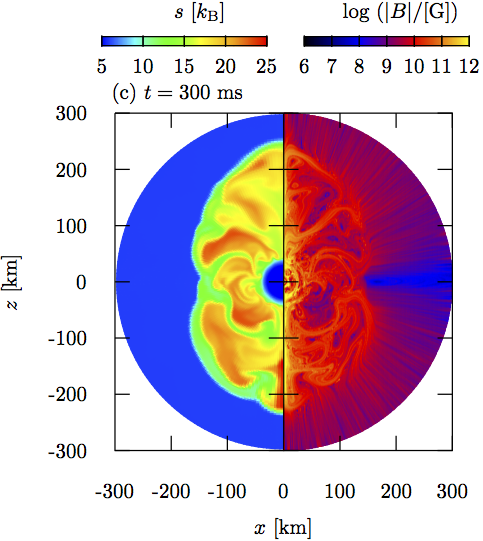}}}
\scalebox{0.9}{{\includegraphics{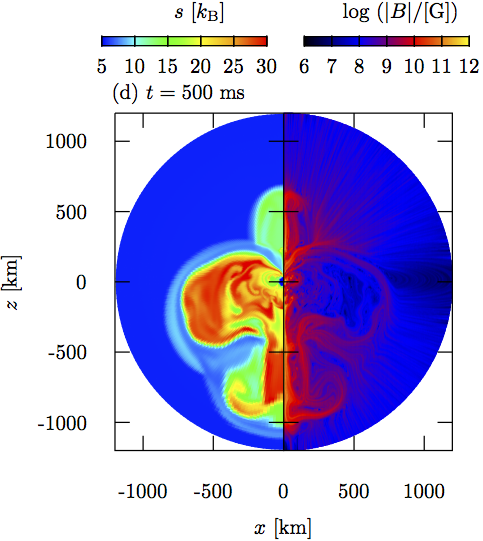}}}
\caption{Time evolution of entropy per baryon ($x<0$)
and magnetic field ($x>0$) for fiducial model (s27.0B10).
Panels (a), (b), (c) and (d) correspond to $t_{\rm pb}=100$, $200$,
$300$ and $500$ ms (final time), respectively. Note that $t_{\rm pb}$
denotes the postbounce time.}
\label{fig1}
\end{center}
\end{figure}

\section{Results} \label{results}
We first describe overall evolution of the magnetized and non-rotating stellar
core for our fiducial model (s27.0B10) in Section~\ref{overall evolution}.
Then in the subsequent sections, we move on to present results focusing on
the impact of the initial magnetic field strength on the postbounce evolution.
The progenitor dependence of the shock evolution is presented in
Section~\ref{progenitor dependence}.

\subsection{Overall evolution of non-rotating and magnetized core-collapse model of a $27 M_{\odot}$ star } \label{overall evolution}

Fig.~\ref{fig1} shows the temporal evolution of the spatial distribution of
the entropy per baryon and magnetic field for the fiducial model (s27.0B10).
The 2D color map of the entropy per baryon is illustrated in the negative
region of $x$ ($x<0$). The structure of magnetic field lines is drawn by a 
line integral convolution method \citep{Cabral93} in the positive region of
$x$ ($x>0$). The color depicts the strength of the magnetic field. Panel~(a),
~(b), ~(c) and ~(d) correspond to the time $t_{\rm pb} =100$, $200$, $300$
and $500$ ms after bounce, respectively. Hereafter $t_{\rm pb}$ denotes
the postbounce time.

\begin{figure}
\begin{center}
\scalebox{0.9}{{\includegraphics{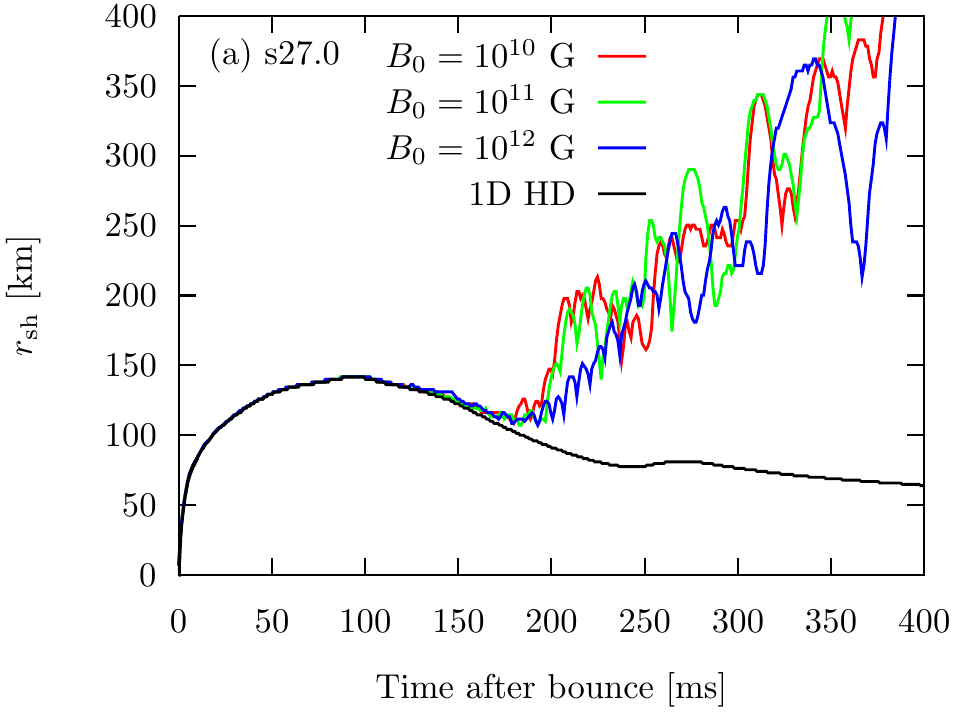}}}
\scalebox{0.9}{{\includegraphics{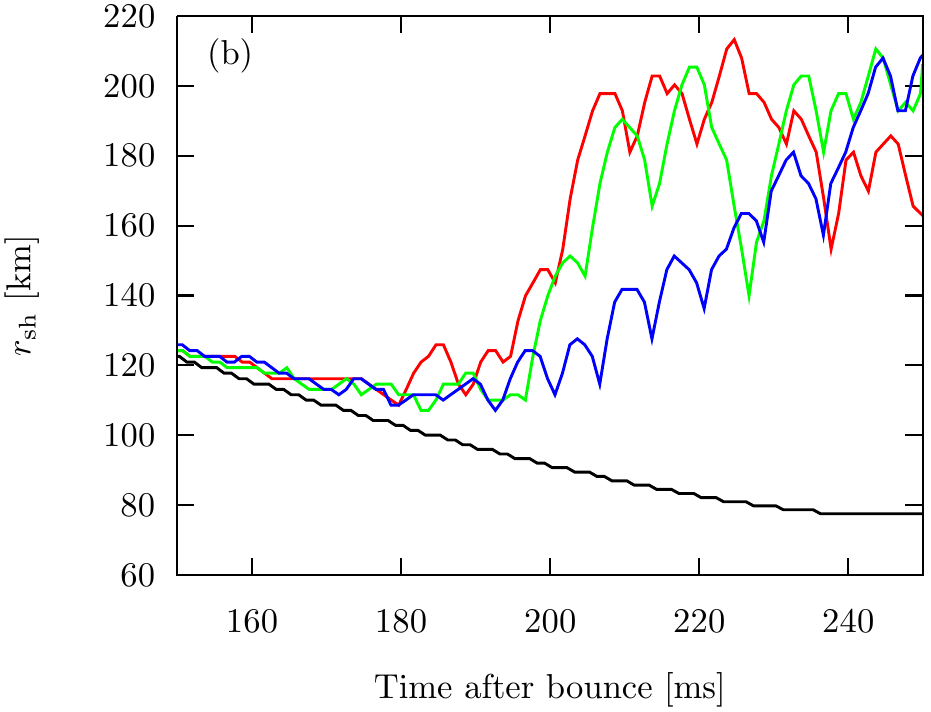}}}
\scalebox{0.9}{{\includegraphics{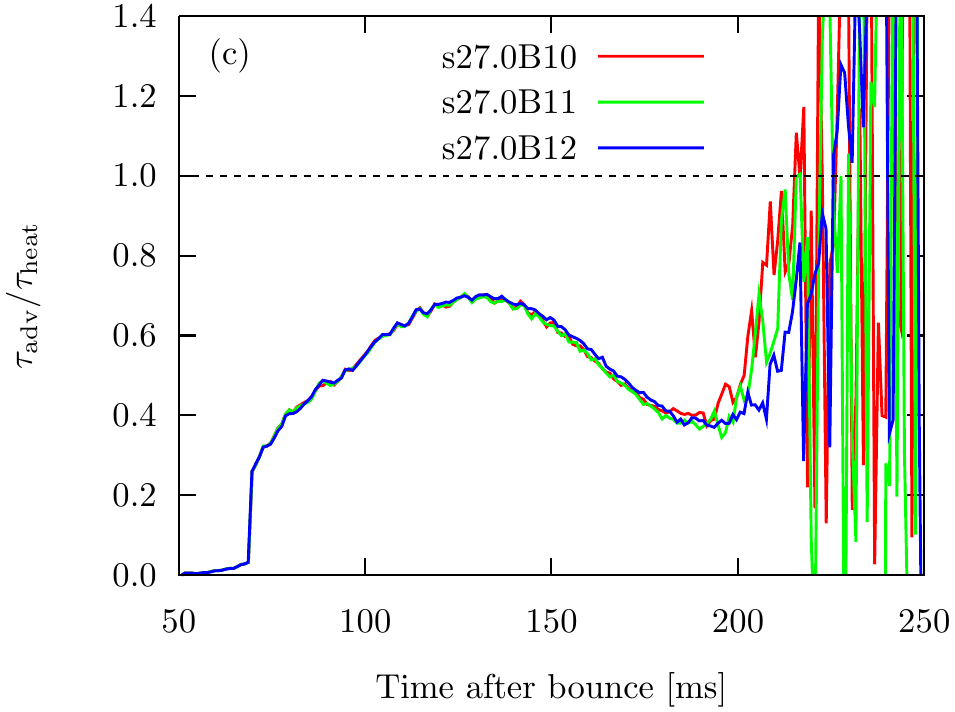}}}
\scalebox{0.9}{{\includegraphics{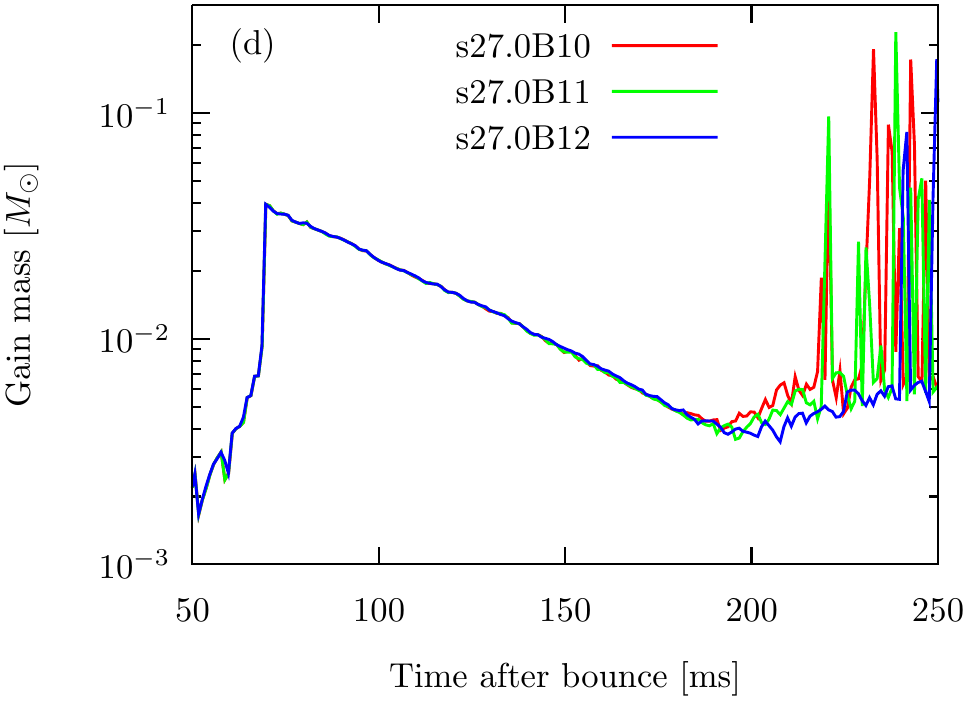}}}
\caption{Time evolution of shock radius (top panels),
$\tau_{\rm adv}/\tau_{\rm heat}$ (bottom left panel) and
gain mass (bottom right panel) for s$27.0$ $M_\odot$ progenitor.
Red, green and blue lines in each panel are the cases for $B_0=10^{10}$,
$10^{11}$ and $10^{12}$ G, respectively. Black lines in top panels show
the evolution of the 1D HD simulation for the $27.0$ $M_\odot$ model.}
\label{fig2}
\end{center}
\end{figure}

The core bounce occurs after $\sim 200$ ms (i.e. $t_{\rm pb}$ = 0) after
the start of the simulation, leading to the shock formation at the radius
of $\sim 20$ km. The bounce shock stalls at $r \sim 140$ km around
$t_{\rm pb} =100$ ms, and then turns into the  standing shock (see also,
the top left panel of Fig.~\ref{fig2}). When the shock stalls, the structure
of the magnetic field lines is like a split monopole as shown in the
right-half panel of Fig.~\ref{fig1}a. Before the shock stall ($t_{\rm pb}
\lesssim 100$ ms), the flow is almost restricted in radial direction.
The split-monopole like configuration is made because the magnetic field is 
"frozen-in" with respect to the matter motion. The electric resistivity of
the magnetic field is so small that it is disregarded in this work, which
can be well justified in the CCSN environment \citep{Sawai13r}. The initial
vector potential (equation \ref{eq: initial magnetic field}) gives magnetic
loops on the equatorial region at around $r \sim 1000$ km. These magnetic
loops also gravitationally collapse (dragged by matter infall) and are shown
on the equatorial plane ($x \gtrsim 30$ km and $z$ = 0) in Fig.~\ref{fig1}a.
The center of loops is located at around $x \sim 45$ km and seen as a small
blueish region.

As the (maximum) shock radius starts to gradually shrink after $t_{\rm pb}
\gtrsim 100$ ms (e.g. Fig.~\ref{fig2}a), it gradually deviates from the shock
trajectory of the corresponding 1D model (black solid line in Fig.~\ref{fig2}a).
This marks the growth of non-spherical motions in the postshock region.
One can clearly observe the deformation of the shock in the left-half panel of
Fig.~\ref{fig1}b at $t_{\rm pb}$ = 200 ms. In Fig. \ref{fig1}b, one can also
see the penetration of the magnetic field lines (thin red curves in the
right-half panel) into the postshock region (high entropy region in the
left-half panel), which makes the field configuration much more complicated
than that outside the shock. In our ideal MHD simulations, the field
amplification in the postshock region occurs due to compression and stretching
of the magnetic field, which is governed by the non-radial matter motions.
Note in our 2D models that we do not attempt to differentiate the origin of
the "non-radial" motions either originating predominantly from the SASI or
neutrino-driven convection because the SASI is liable to be overestimated in
2D compared to 3D simulations (e.g. \citealt{Hanke12,Hanke13,Rodrigo14}). 
 
Fig.\ref{fig1}c shows a snapshot after the shock revival ($t_{\rm pb} =$
300 ms, see also Fig. \ref{fig2}a). The low-mode deformation of the shock and
the formation of the high entropy region (colored by red in the entropy plot)
is a common feature of 2D neutrino-driven explosion models. The highly
aspherical shock (Fig. \ref{fig1}d) continues to propagate strongly toward
the south pole up to a radius of $\sim 1000$ km at the final calculation time
($t_{\rm fin} \sim 500$ ms) for this model. The magnetic field configuration is
similar to the blast morphology as in the previous snapshots.

\subsection{Neutrino-driven explosion of MHD core collapse} \label{neutrino-dirven explosion}

Fig.~\ref{fig2} summarizes the time evolution of the maximum shock radius
(top panels), the ratio of the advection timescale to the neutrino-heating
timescale, $\tau_{\rm adv}/\tau_{\rm heat}$, (bottom left panel) and gain
mass (bottom right panel) for the $27.0$ $M_\odot$ model. Red, green and
blue lines in each panel are the cases for $B_0=10^{10}$, $10^{11}$ and
$10^{12}$ G, respectively.

In Fig.~\ref{fig2}a, a black line, as already mentioned, denotes the shock
evolution of the 1D HD model of the $27.0$ $M_\odot$ star, which is shown
as a reference. The shock radius in 1D maximally reaches to $\sim 140$ km
at $t_{\rm pb} \sim 100$ ms, then continues to contract during the simulation
time. On the other hand, the shock revival occurs in the 2D MHD models
regardless of the strength of the initial magnetic field. Looking at the
panel very carefully, one might notice an interesting tendency about the
slight delay of the onset of the shock revival between the three MHD models. 

To show this clearly, we make a comparison in Fig.~\ref{fig2}b of the shock
evolution only close to the shock revival time (between $t_{\rm pb} =150$ ms
and $t_{\rm pb} =250$ ms). The shock revival time is indeed delayed for the
strong initial field model (blue curve) comparing to the weak initial field
model (red curve).

Fig.~\ref{fig2}c shows the evolution of the timescale ratio,
$\tau_{\rm adv}/\tau_{\rm heat}$. Following \citet{Summa16}, we estimate the
advection timescale as
\begin{eqnarray}
\tau_{\rm adv} = \frac{M_{\rm g}}{\dot{M}} \; ,
\end{eqnarray}
where $M_{\rm g}$ is the mass enclosed in the gain layer (gain mass) and 
$\dot{M}$ is the mass-accretion rate through the shock. The neutrino-heating
timescale is defined by
\begin{eqnarray}
\tau_{\rm heat} = \frac{|E_{\rm tot,g}|}{\dot{Q}_{\rm heat}} \; ,
\end{eqnarray}
where $|E_{\rm tot, g}|$ is the total energy of the material in the gain layer
and $\dot{Q}_{\rm heat}$ is the neutrino-heating rate in this region. Since
the residency time of matter in the gain region is related to the exposure
of the material to neutrino heating, $\tau_{\rm adv}/\tau_{\rm heat} \gtrsim 1$
is a necessary condition for the onset of the shock revival (e.g. \citealt{Buras06}). 
As shown in Fig.~\ref{fig2}c, $\tau_{\rm adv} /\tau_{\rm heat}$ rises toward
unity rapidly at around $t_{\rm pb} \sim 200$ ms. It is noted that this shock
revival timescale is consistent with that by \citet{Hanke13} who conducted the
2D model using the same progenitor ($27 M_{\odot}$ star of \citealt{Woosley02})
with more elaborate neutrino transport scheme.

In Fig.~\ref{fig2}c, it is important to point out that the growth rate of the
timescale ratio (before exceeding unity) is highest for the weakly magnetized
model (red line), which is followed in order by the moderately magnetized model
(green line) and the strongly magnetized model (blue line). This feature is
closely linked to the shock evolution after $t_{\rm pb} \sim 200$ ms, namely,
the onset of the shock revival and the subsequent runaway shock expansion is
delayed for the strongly magnetized models. This indicates that the neutrino
heating mainly contributes to the runaway shock expansion as shown in
Fig.~\ref{fig2}a. And the magnetic field secondary affects the shock evolution.
The same tendency is also observed in Fig.~\ref{fig2}d, which compares the
evolution of the gain mass at around $t_{\rm pb} \sim 200$ ms. 

\begin{figure}
\begin{center}
\scalebox{0.9}{{\includegraphics{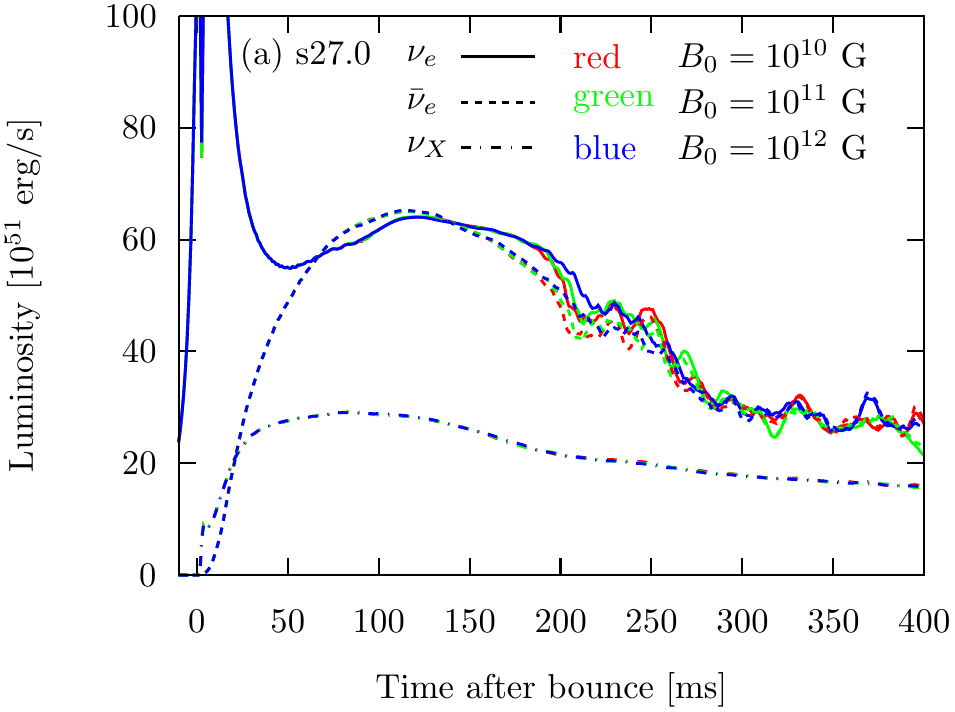}}}
\scalebox{0.9}{{\includegraphics{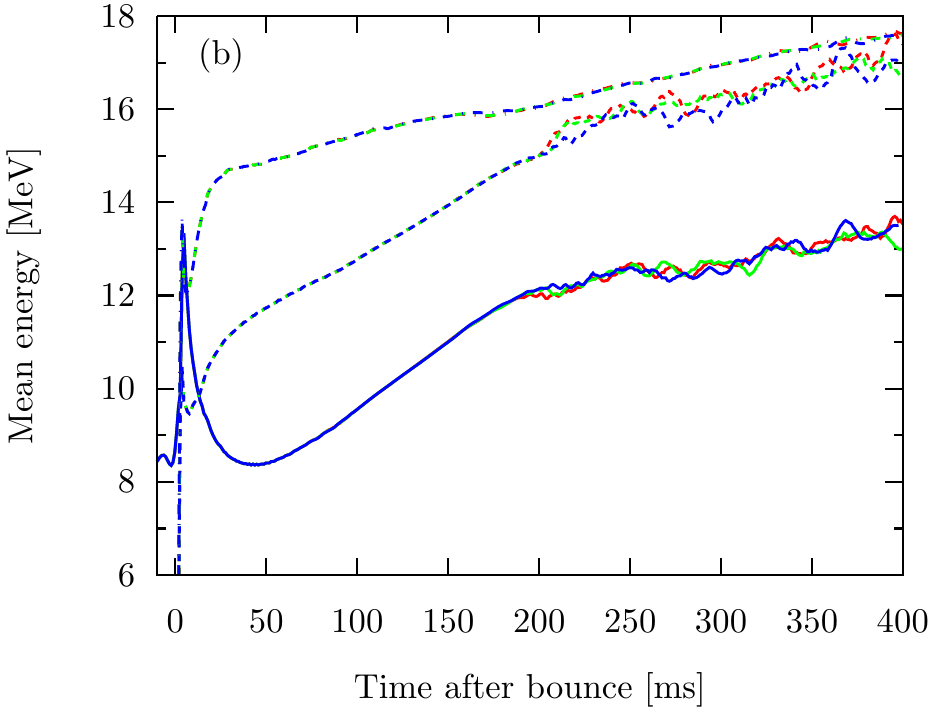}}}
\caption{Panel (a): temporal evolution of neutrino luminosity for fiducial
progenitor model (s27.0). Panel (b): time evolution of mean energy of
neutrinos for fiducial progenitor model (s27.0). Line types and colors
have the same meanings as those in panel (a).}
\label{fig3}
\end{center}
\end{figure}

Fig.~\ref{fig3} compares the time evolution of (a) neutrino luminosity
and (b) mean energy of neutrinos between the model s27.0 series. Solid,
dashed and dash-dot lines correspond to $\nu_e$, $\bar{\nu}_e$ and $\nu_X$,
respectively. Red, green and blue lines are the cases for $B_0=10^{10}$,
$10^{11}$ and $10^{12}$ G, respectively. The evolution of $\nu_e$ and
$\bar{\nu}_e$ with  different magnetic fields in the luminosity and mean
energy are almost identical up to $200$ ms and represented as blue solid
and dashed lines, respectively. On the other hand, the evolution of
$\nu_X$ with different  magnetic fields almost overlaps during the whole
calculation time. Given the results mentioned above, it may not be so
surprising that the  initial magnetic fields have little impact on the
luminosities and mean energies. It is also noted that they are in good
agreement with those in \citet{Summa16} who have done a systematic 2D
simulations covering a wide range of the (non-magnetized) progenitor models.
Regardless of the difference in the neutrino transport scheme, the peak of the
$\nu_e$/$\bar{\nu}_e$  luminosity (for the same progenitor) is $\sim 6 \times
10^{52}$ erg/s at around $t_{\rm pb} \sim $ 100 ms, which is consistent with
our results. After the onset of shock revival ($t_{\rm pb} \gtrsim 200$ ms),
the mean neutrino energy is in the range of 12 $\sim  14$ MeV for $\nu_e$ and
15 $\sim  17$ MeV for $\bar{\nu}_e$ in our model, which nicely matches with
\citet{Summa16}, although our $\bar{\nu}_e$ mean energy is $\sim 8\%$ higher
than \citet{Summa16} at our final simulation time.

Although there is no significant impact of the initial magnetic field strength
on the neutrino properties (Fig.~\ref{fig3}), we did witness the difference in
the evolution of $\tau_{\rm adv}/\tau_{\rm heat}$ (Fig.~\ref{fig2}c) for models
with stronger initial magnetic fields. This suggests that the stronger initial
field affects the advection timescale predominantly than the neutrino heating
timescale. In what follows, we explore how the strong initial field could
affect the development of the non-radial matter motions in the postshock
region, leading to the delayed onset of the shock revival.

\begin{figure}
\begin{center}
\scalebox{0.85}{{\includegraphics{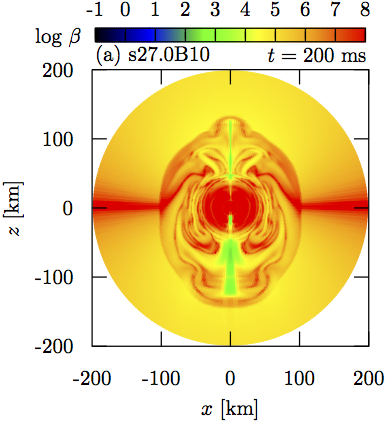}}}
\scalebox{0.85}{{\includegraphics{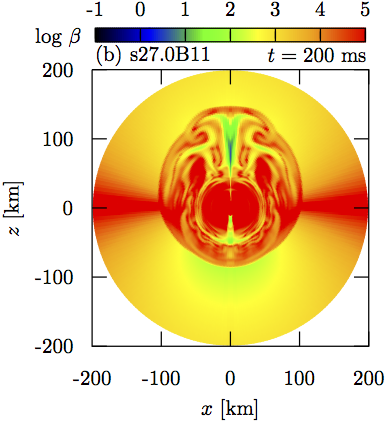}}}
\scalebox{0.85}{{\includegraphics{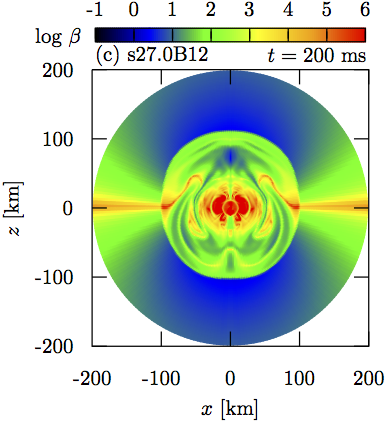}}}
\caption{2D distribution of plasma $\beta$ for fiducial progenitor
model (s27.0) at $t_{\rm pb} = 200$ ms. Panels (a), (b) and (c) correspond
to the cases with $B_0=10^{10}$, $10^{11}$ and $10^{12}$ G, respectively.}
\label{fig4}
\end{center}
\end{figure}

One may wonder whether the amplified magnetic field in the postshock region
could assist the explosion as reported in \citet{Obergaulinger14}. The plasma
$\beta$, the ratio of the thermal pressure to the magnetic pressure, is a good
indicator to check this possibility. Fig.~\ref{fig4} is a comparison of 2D
spatial distribution of the plasma $\beta$ at $t_{\rm pb} =200$ ms (e.g. close
to the shock revival time, see Fig.~\ref{fig2}b) for the $27.0$ $M_\odot$
models with different initial magnetic fields. It is shown that the plasma
$\beta$ behind the shock is much larger than unity ($\gtrsim 10^{1} \sim 10^{5}$),
meaning that the shock revival in our models is predominantly driven by
neutrino heating. This is absolutely not the case in the MHD explosion in
the context of rapidly rotating and strongly magnetized cores (e.g.
\citealt{Kuroda20}), where the plasma $\beta$ of $\sim$ unity is often 
achieved. This is because the rapidly rotating PNS with strong differential 
rotation at the surface results in the magnetic field winding and the 
subsequent increase of the magnetic pressure. The increased magnetic
pressure derives the shock expansion toward the polar directions.
Furthermore it is important to note that the neutrino-driven shock revival
is obtained at $t_{\rm pb} \sim 200$ ms for our weakly magnetized model
(s27.0B10). This is in stark contrast with \citet{Obergaulinger14} who obtained
a very {\it late} onset of explosion at $t_{\rm pb} \sim 700$ ms for a 15
$M_{\odot}$ progenitor employed in the work, though assuming the same magnetic
field strength  ($10^{10}$ G). More detailed comparison between our models and
the previous studies is presented in Section~\ref{progenitor dependence}.

\subsection{Magnetic field dependence on non-radial motion and development of turbulence} \label{magnetic dependence}

\begin{figure}
\begin{center}
\scalebox{0.9}{{\includegraphics{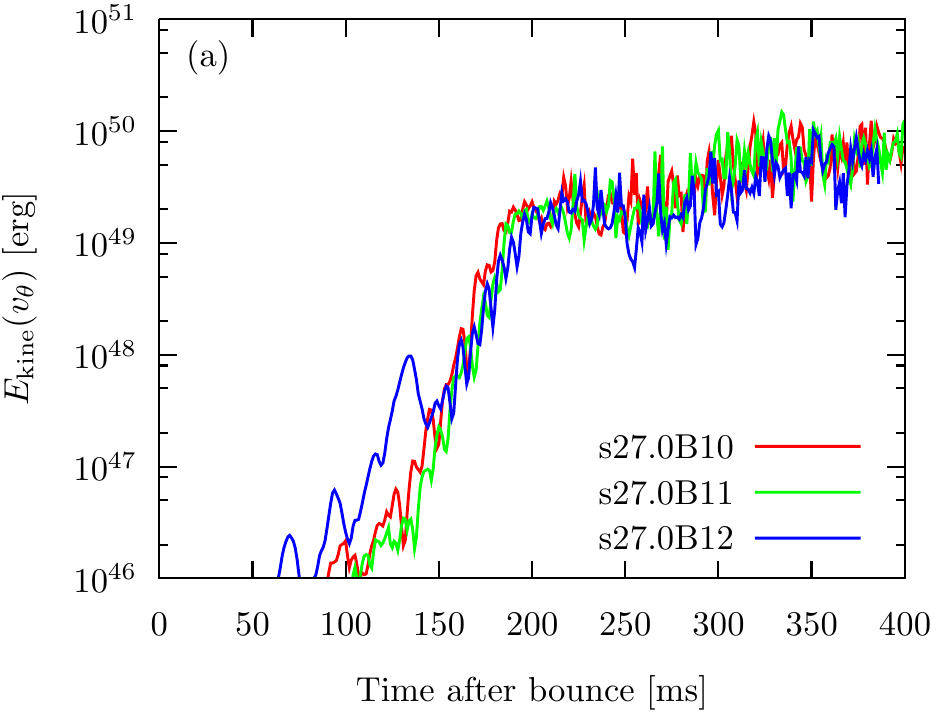}}}
\scalebox{0.9}{{\includegraphics{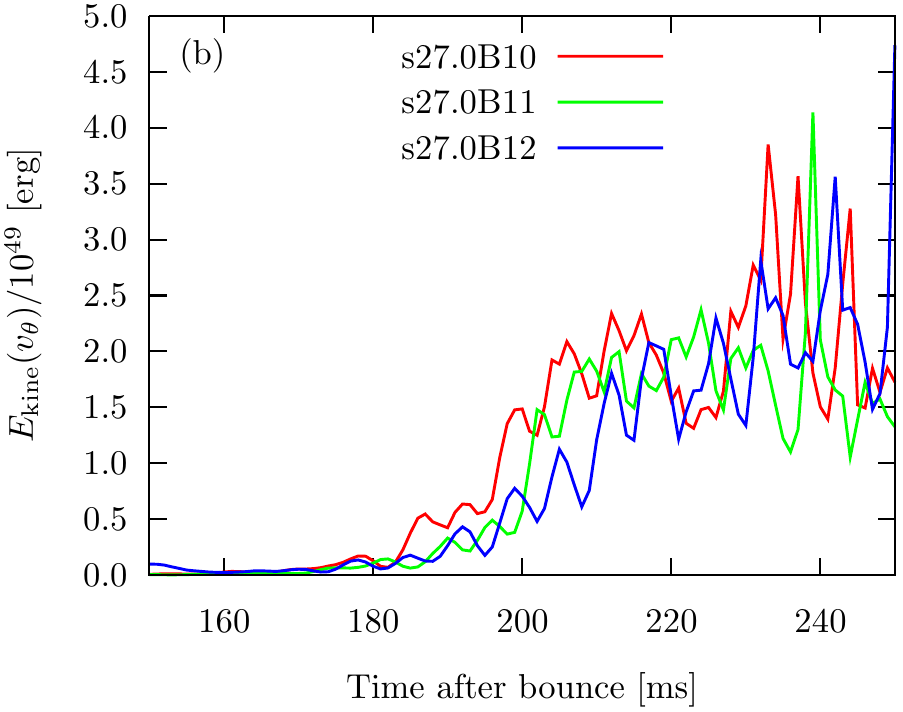}}}
\scalebox{0.9}{{\includegraphics{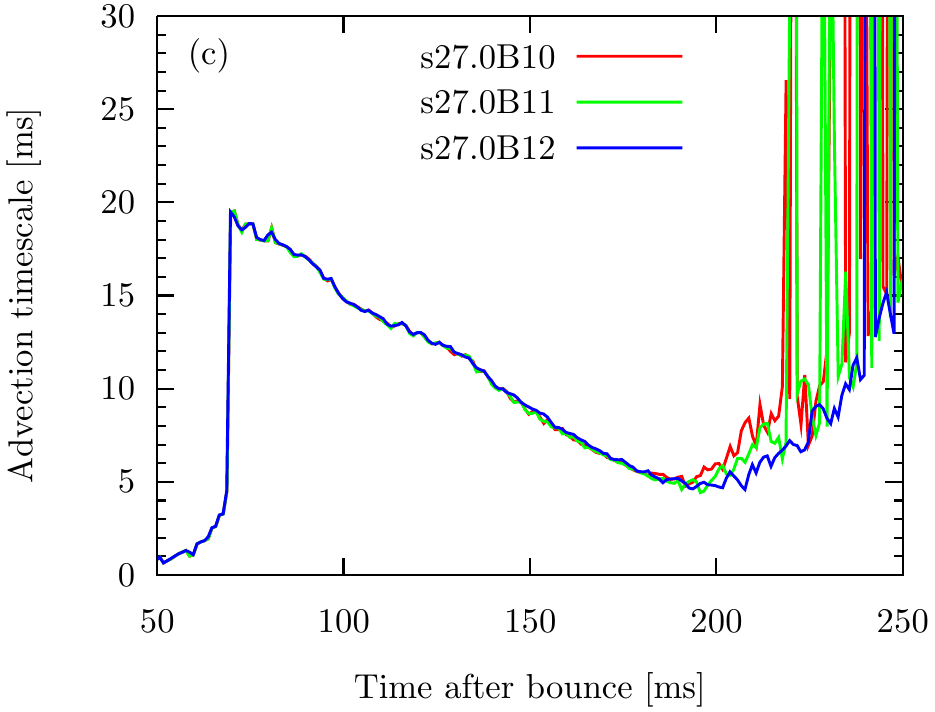}}}
\scalebox{0.9}{{\includegraphics{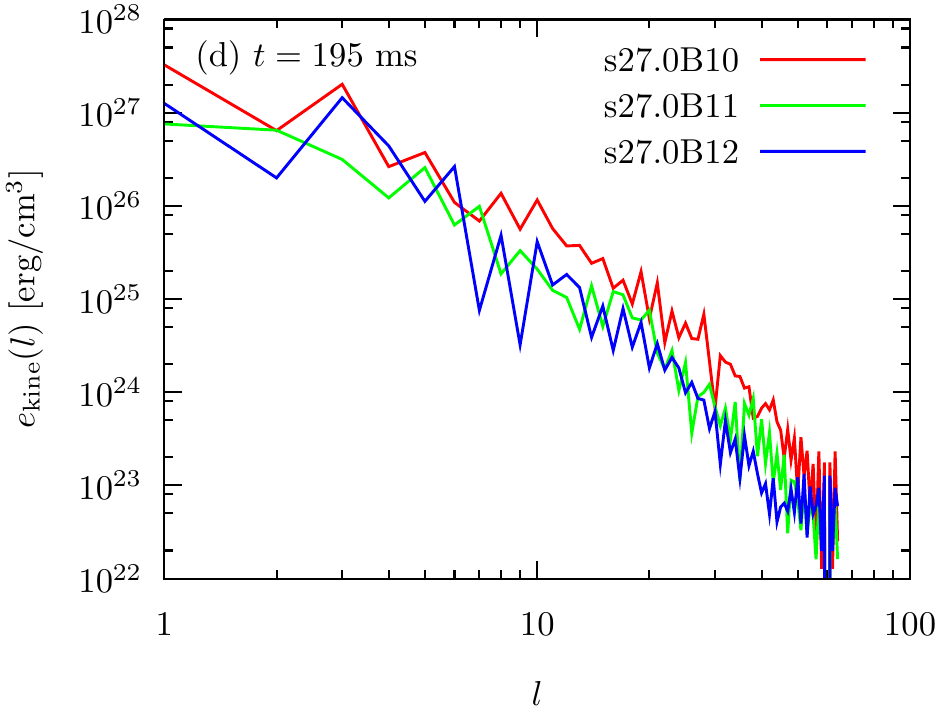}}}
\caption{Temporal evolution of kinetic energy (top panels), time
evolution of advection timescale (bottom left panel) and spectrum of
kinetic energy at $t=204$ ms (bottom right panel) in the gain region
for the fiducial progenitor model (s27.0). In order to focus on the
convective motion, only $\theta$-component of velocity is taken into
account to compute the kinetic energy. The energy spectrum is defined
by equation (\ref{eq: turbulent energy spectrum}). Red, green and blue
lines in each panel are the cases for $B_0=10^{10}$, $10^{11}$ and
$10^{12}$ G, respectively.}
\label{fig5}
\end{center}
\end{figure}

In the previous subsections, we have shown that the neutrino heating plays
a dominant role in triggering the explosion of our 27 $M_{\odot}$ models,
whereas the magnetic field plays a secondary role, namely, to delay the
explosion onset.  We now proceed to clarify the reason by focusing on the role
of the magnetic field on the non-radial matter motions and turbulence in the
postshock region.

Similar to Fig.~\ref{fig2}, but Fig.~\ref{fig5}a, ~\ref{fig5}b and ~\ref{fig5}c
show the time evolution of the lateral kinetic energy (top panels), and the
advection timescale in the gain region for the model series of s27.0,
respectively. In Fig.~\ref{fig5}a, the contribution from the lateral
($\theta$)-component of the velocity ($v_{\theta}$) is taken into account as
a measure to quantify the vigor of the non-radial motions and turbulence in
the gain region. The entire evolution of the lateral kinetic energy is shown
in Fig.~\ref{fig5}a (log-scale in the $y$-axis), whereas Fig.~\ref{fig5}b
focuses on the time around the shock revival ($t_{\rm pb} \sim 200$ ms)
(linear scale in the $y$-axis). From Fig.~\ref{fig5}a, one can see that the
lateral kinetic energies firstly increase exponentially before the shock
revival ($t_{\rm pb} \sim 200$ ms), and then reach asymptotically to
$\sim 10^{50}$ erg toward the final simulation time regardless of the
different initial field strength.

Looking more closely at the linear phase ($t_{\rm pb} \lesssim 200$ ms),
Fig.~\ref{fig5}b depicts that the growth of the kinetic energy is fastest
(biggest) for model s27.0B10 (red line) compared to the more strongly
magnetized models (green and blue lines). This feature, as previously
identified in the 3D MHD simulations of \citet{Endeve12} (but with more
idealized setting), is also consistent with the earlier onset of the shock
revival as seen in Fig.~\ref{fig2}b. 

Since the neutrino heating timescale is similar among the s27.0 models (e.g.
Section~\ref{neutrino-dirven explosion}), the difference of the advection
timescale in the gain region should play a key role of the explosion onset,
i.e. the longer the better. As shown in Fig.~\ref{fig5}c, the advection
timescale of the weakly magnetized model (red line) is actually longer than
that of the strongly magnetized model (blue line) around the explosion onset
($t_{\rm pb} \sim 200$ ms). The stronger nonradial motions obtained in the
weakly magnetized model (red line) (Figs.~\ref{fig5}a,b) are consistent with
the longer advection time of the material in the gain layer as seen in
Fig.~\ref{fig5}c. Likewise, these results are in favour of explaining
the delayed onset of the shock revival of the strongly magnetized models.

In order to investigate the role of the initial magnetic field on the turbulent
activity in the postshock region, we compute the turbulent energy spectrum,
$e_{\rm kine}(l)$. Following \citet{Hanke12}, it is defined as,
\begin{eqnarray}
e_{\rm kine}(l)=\sum_{m=0} \biggl | \int_{\Omega} Y_{lm}^{*}(\theta,\phi) 
\sqrt{\rho}v_{\theta}(r,\theta,\phi)d\Omega \biggr |^2 \; , \label{eq: turbulent energy spectrum}
\end{eqnarray}
where $Y_{lm}$ is the spherical harmonics of degree $l$ and $m$, and
$\Omega$ is a solid angle. Note in our 2D simulations, $m=0$ is only
considered in equation (\ref{eq: turbulent energy spectrum}). The turbulent
energy spectra in Fig.~\ref{fig5}d are evaluated at a fixed radius ($r=100$ km)
in the postshock region and at around the explosion onset. They are time
averages at the center of $t_{\rm pb}=195$ ms.

Fig.~\ref{fig5}d clearly shows that the energy density of higher-order modes
($l \gtrsim 10$) is bigger for the weaker magnetic field case (red line) than
that for the stronger field case (blue line), although the energy density of
the smaller-order modes is comparable in all the three cases. This implies that
the strong magnetic field, most likely due to the magnetic tension, prevents
the growth of the turbulent motions  down to small scales (at larger $l$).
The suppression of the turbulent energy at larger $l$ is reconciled with the
slow increase of the non-radial kinetic energy for the strongly magnetized
model as shown in  Fig.~\ref{fig5}b. Again this is consistent with the delayed
onset of the shock revival.

\begin{figure}
\begin{center}
\scalebox{0.9}{{\includegraphics{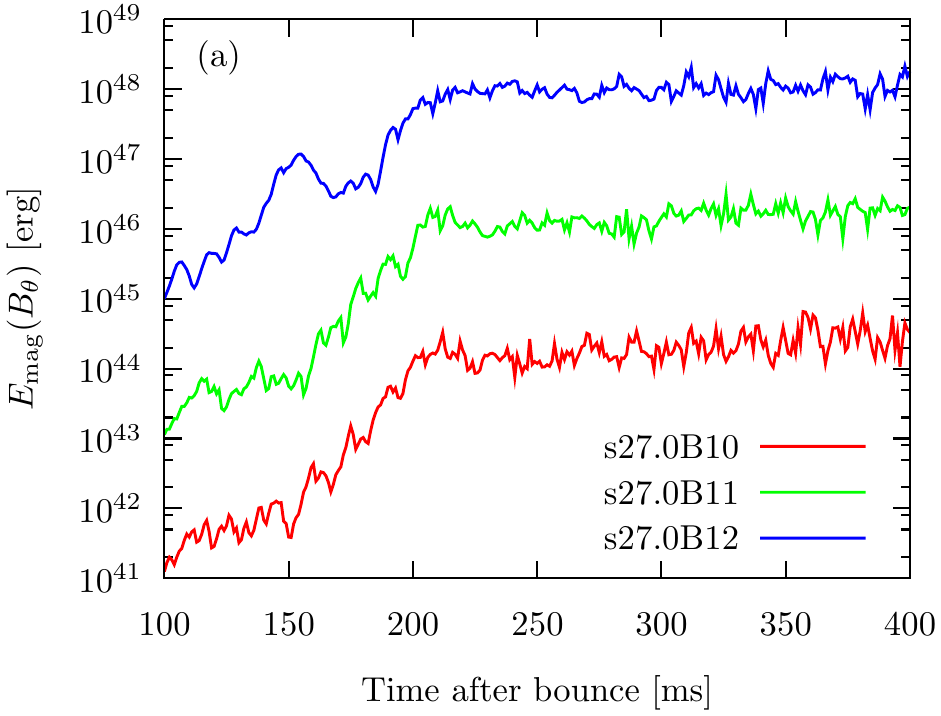}}}
\scalebox{0.9}{{\includegraphics{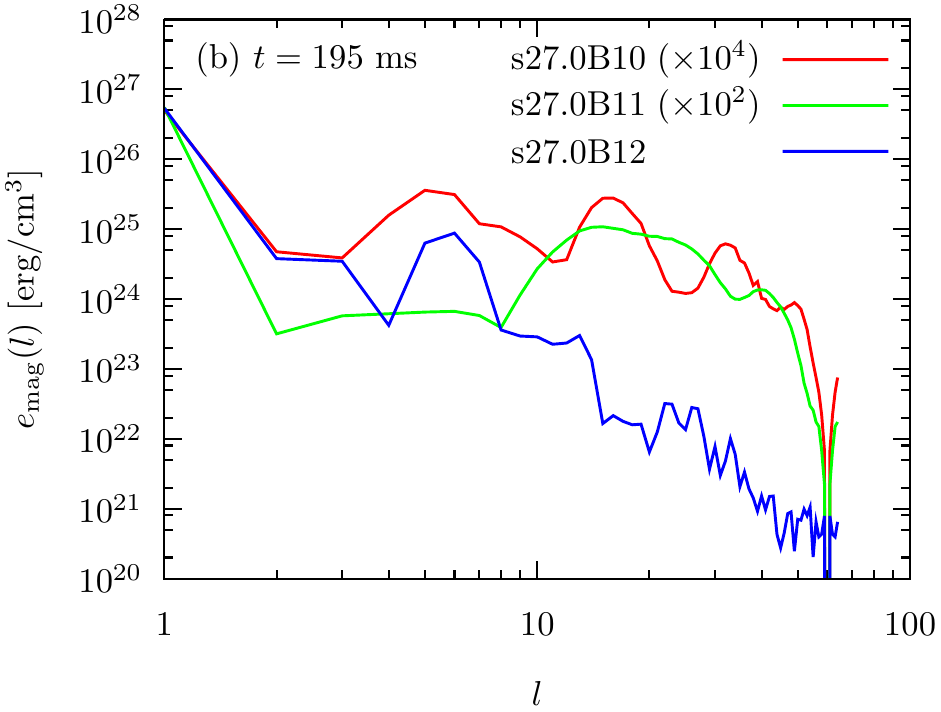}}}
\caption{Panel (a): temporal evolution of magnetic energy (only
$\theta$-componet) in gain region for fiducial progenitor model (s27.0).
Red, green and blue lines are the cases for $B_0=10^{10}$, $10^{11}$
and $10^{12}$ G, respectively. Panel (b): spectrum of magnetic energy
in gain region defined by equation (\ref{eq: magnetic turbulence spectrum})
for fiducial progenitor model at $t=197$ ms. The meanings of colors for
lines are the same as those in panel (a). In order to normalize the spectra
with different field strength, those for the case with $B_0=10^{10}$ and
$10^{11}$ G are multiplied by a factor of $10^4$ and $10^2$, respectively.}
\label{fig6}
\end{center}
\end{figure}

Similar to Fig.~\ref{fig5}a and~\ref{fig5}d, but Fig.~\ref{fig6} shows the
evolution of the lateral magnetic energy in the gain region (left panel) and
the corresponding time-averaged energy spectra of the magnetic turbulence at
$t_{\rm pb} = 195$ ms (right panel). Fig.~\ref{fig6}a shows that the lateral
magnetic energy is exponentially amplified up to the explosion onset
($t_{\rm pb} \sim 200$ ms) in all the three models. The exponential growth
terminates when the shock revival initiates ($t_{\rm pb} \gtrsim 200$ ms).
In the non-linear phase ($t_{\rm pb} \gtrsim 200$ ms), the lateral magnetic
energies are shown to be almost kept constant with time, whose strength is
bigger for the strongly magnetized model. It is noted that the saturated
values of model s27.0B12 ($\sim 10^{48}$ erg, blue line), s27.0B11
($\sim 10^{46}$ erg, green line), and s27.0B10 ($\sim 10^{44}$ erg, green line)
differ by the two orders-of-magnitudes, respectively, which is proportional to
the square of the initial magnetic energy (i.e. $\propto B_0^2$). 

Since the magnetic pressure/force does not play a crucial role in the shock
revival (see Fig.~\ref{fig4} and the explanation in the last paragraph of
Section~\ref{neutrino-dirven explosion}), the magnetic field can be amplified
by the compression and stretching due to the non-radial fluid motions. 
Actually, by comparing Fig.~\ref{fig5}a and Fig.~\ref{fig6}a, one can see that
the magnetic energy inside the gain region is less than the kinetic energy in
all the three models. The ratio of the magnetic energy to the kinetic energy,
$E_{\rm mag}(B_{\theta})/E_{\rm kine}(v_{\theta})$, in models s27.0B10,
s27.0B11 and s27.0B12 at around $t_{\rm pb} \sim 200$ ms are $10^{-5}$,
$10^{-3}$ and $10^{-1}$, respectively, whereas the kinetic energy at the time
is almost $10^{49}$ erg similar to all the three models. The shock revival
occurs before the magnetic energy in the gain layer reaches to the
equipartition with the kinetic energy. We speculate if the explosion were
much more delayed like in \citet{Obergaulinger14}, the magnetic field
amplification could continue until it becomes sufficiently high (such as
the equipartition level) enough to assist the shock revival.

To quantify the vigor of turbulence of the magnetic field, we estimate
the spectrum of the lateral magnetic field as in equation
(\ref{eq: turbulent energy spectrum}). Plotted in Fig.~\ref{fig6}b is
\begin{eqnarray}
e_{\rm mag}(l)=\sum_{m=0} \biggl | \int_{\Omega} Y_{lm}^{*}(\theta,\phi)
B_{\theta}(r,\theta,\phi)d\Omega \biggr |^2 \;. \label{eq: magnetic turbulence spectrum}
\end{eqnarray}
In order to normalize the spectra with the different initial field strength,
the spectra for the case with $B_0=10^{10}$ and $10^{11}$ G are multiplied
by a factor of $10^4$ and $10^2$, respectively. This is reasonable because
the magnetic energy is proportional to $B^2$ and that of $B_0=10^{10}$ and
$10^{11}$ G are $10^4$ and $10^2$ times smaller than that of $B_0=10^{12}$ G,
respectively.

From Fig.~\ref{fig6}b, one can see that the energy spectrum of the magnetic
turbulence typically decreases with $l$, i.e. the magnetic energy is mainly
stored in a large-scale field. The slope is, however, shallower for the weak
field model (s27.0B10, red line) comparing with the strong field models (green
and blue lines). This indicates that a small scale (turbulent) magnetic field
can be more preferentially developed for the weak field model. The relative
excess of the lateral (turbulent) magnetic energy at larger $l$ is consistent
with the excess of the lateral (turbulent) kinetic energy  as shown in 
Fig.~\ref{fig5}d. These results suggest that the strong initial field could
act to suppress the magnetic turbulence in the small scale compared to that
for the weak initial field model.

\subsection{Impact of different progenitor models} \label{progenitor dependence}

\begin{figure}
\begin{center}
\scalebox{0.9}{{\includegraphics{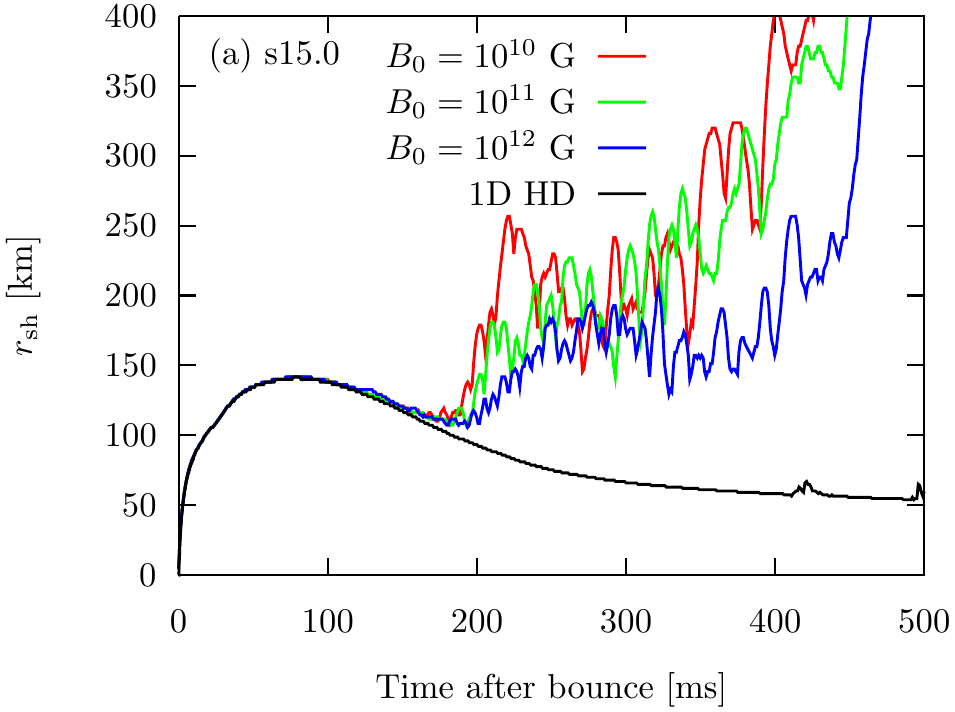}}}
\scalebox{0.9}{{\includegraphics{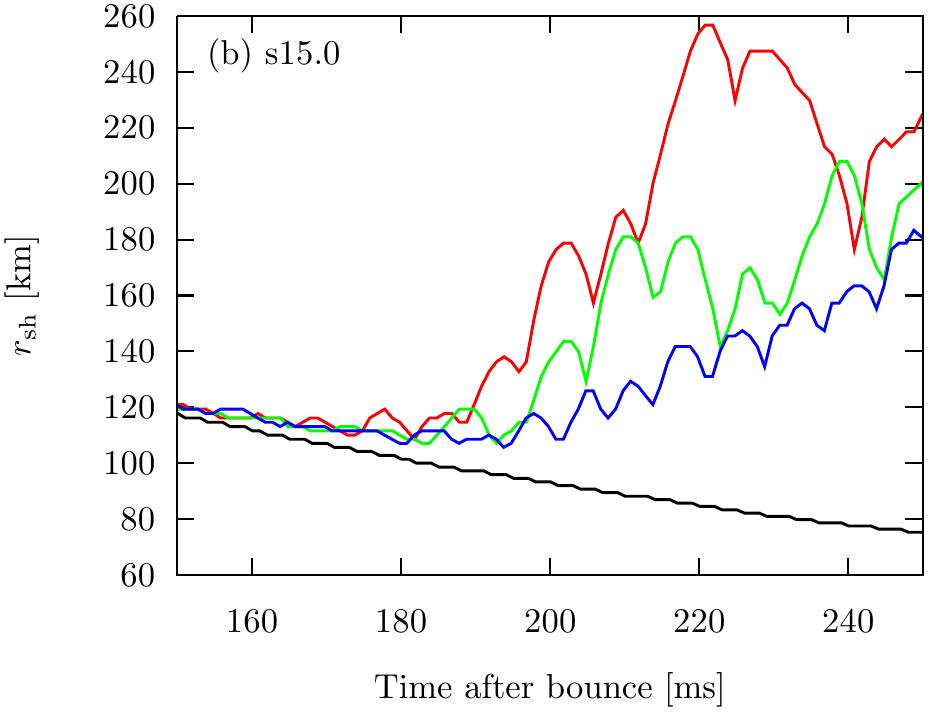}}}
\scalebox{0.9}{{\includegraphics{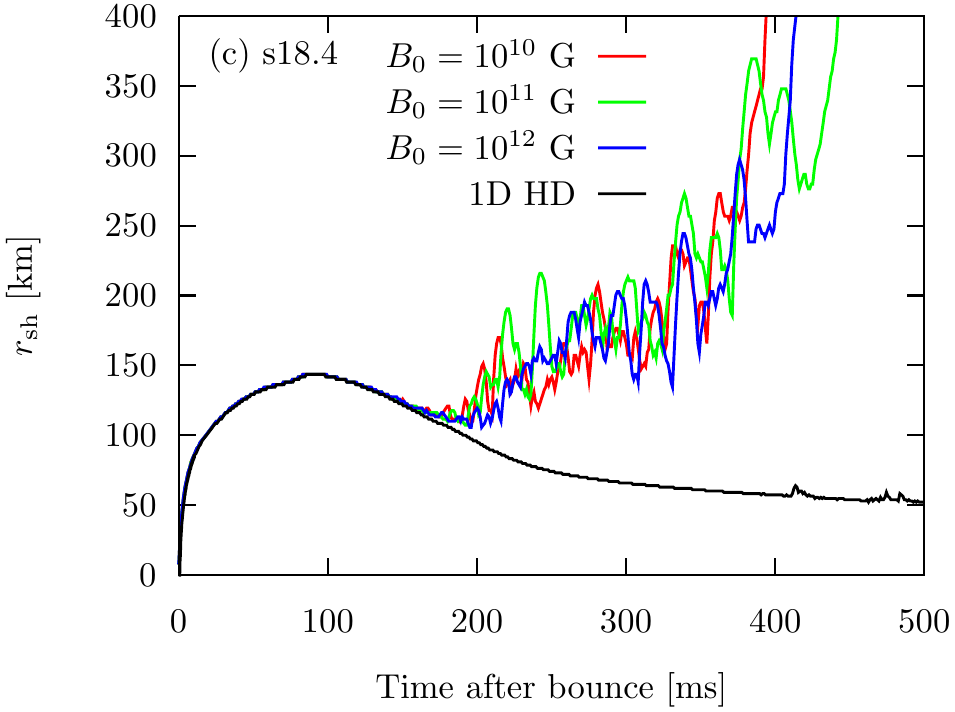}}}
\scalebox{0.9}{{\includegraphics{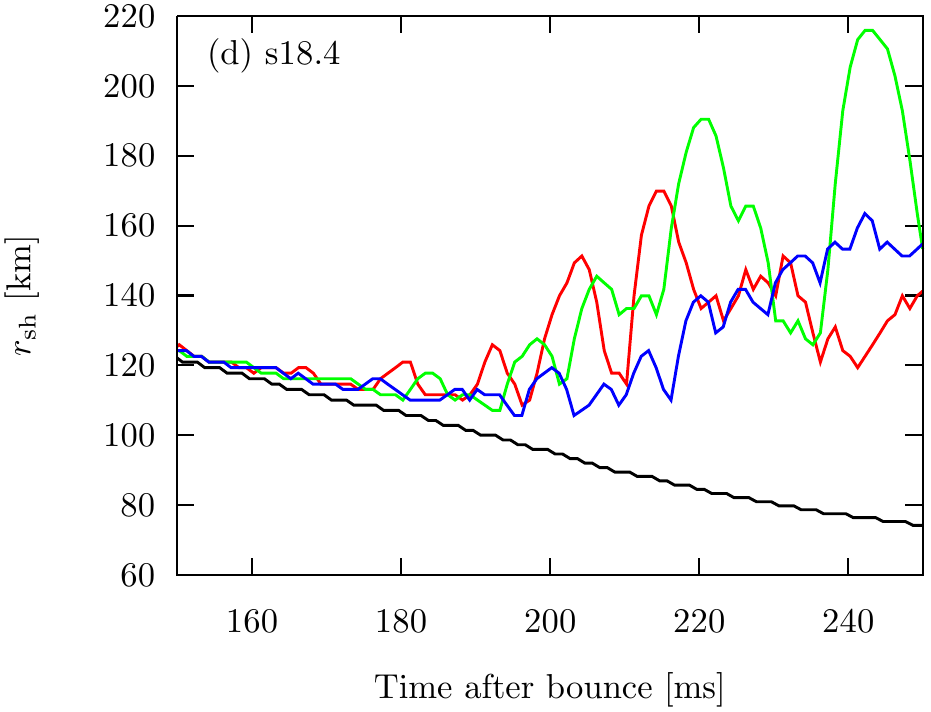}}}
\caption{Evolution of shock radii for models s15.0 and s18.4. Red, green
and blue lines in each panel are the cases for $B_0=10^{10}$, $10^{11}$
and $10^{12}$ G, respectively. Black lines show the evolution of the 1D
calculation (HD) for each model.}
\label{fig7}
\end{center}
\end{figure}

In order to investigate the impact of the initial magnetic field on the
different progenitor models, we present results for the $15.0$ and $18.4$
$M_\odot$ models. Similar to Figs.~\ref{fig2}a and~\ref{fig2}b, but
Fig.~\ref{fig7} shows the evolution of the shock for model $15.0$ $M_\odot$
(top panels) and model $18.4$ $M_\odot$ (bottom panels), respectively.
In both of the models, the shock revival occurs at $t_{\rm pb} \sim 200$ ms
(Fig.~\ref{fig7}) regardless of the difference in the initial magnetic
field strength. Remarkably, the slight delay of the shock revival for the
strongly magnetized models is also obtained (s15.0B12 vs. s15.0B10 and
s18.4B12 vs. s18.4B10). These features are  common to those obtained in
the $27 M_{\odot}$ models as already mentioned in the last section.

The 2D HD simulation of the $18.4$ $M_\odot$ progenitor model was reported in
\cite{Summa16}. They obtained the shock revival at $t_{\rm pb} \sim 520$ ms,
which is  $\sim$ 320 ms later compared to our  counterpart model (s18.4B10).
Given the big difference of neutrino transport scheme between the two codes,
we cannot unambiguously specify the reason of the discrepancy. But already in
the 1D comparison work of \citealt{OConnor18} (e.g. green line of their
Fig.~\ref{fig2}), we can see that our HD code (3DnSNe) leads to a larger shock
radius especially later than $t_{\rm pb} \sim 200\,{\rm  ms}$ relative to
the other codes, leading to the enhanced heating rate in the gain region
$t_{\rm pb} \gtrsim 250\,{\rm  ms}$ (see, green line of their Fig.~\ref{fig5}).
This could be one of the reasons of the early shock revival seen in our $18.4$
$M_{\odot}$ run. Be that as it may, the shock revival time and the neutrino
properties (Figs.~\ref{fig2} and~\ref{fig3}) show a good agreement with
those of the $27 M_{\odot}$ model of \citet{Summa16}. The match may be simply
by chance, and a detailed comparison of 2D  CCSN models from different groups
is apparently needed to clarify these features. By running the ALCAR code for
a given progenitor model (s20 of \citealt{Woosley07}) but with varying
the neutrino opacities and the transport schemes, \citet{Just18} showed in
their detailed, 2D systematic simulations that a simplified treatment in
the neutrino-pair processes makes the onset time of the shock revival
significantly earlier ($\sim 100 - 400$ ms) compared to the case without
such simplification (compare the explosion time of models s20-rbr and
s20-rbr-pp\{1/2/3\} in their Table~1). Since we employ the simplified
prescription (e.g. the assumption of the isotropy and local-thermal-equilibrium
with respect to the heavy-lepton neutrinos), this might be also the reason of
our earlier onset of the explosion.

As already mentioned before, the 2D MHD simulations of the non-rotating cores
of the $15.0$ $M_\odot$ star \citep{Woosley02} were reported in \citet{Obergaulinger14}
with varying the initial field strength (like in this study). In their strong
initial field model ($10^{12}$ G),  they obtained a magnetically assisted
explosion at $t_{\rm pb} \sim 600$ ms, which is about $\sim 200$ ms earlier
than their weak field model ($10^{10}$ G). If the neutrino-driven shock
revival is so much delayed, they pointed out that an equipartition between
the turbulent kinetic and magnetic energy was archived due to the growing
turbulence over the long period of time. This may seem to contradict with our
results, i.e. the slight delay of the explosion onset for the strongly
magnetized model (s15.0B12) as shown in Fig.~\ref{fig7}a and Fig.~\ref{fig7}b.
However, in our models, the shock revival occurs much earlier until the
equipartition could be achieved. The difference of the 2D hydrodynamics, which
could stem from the omission of heavy-lepton neutrinos in \citet{Obergaulinger14}
and also from the difference of the neutrino transport scheme (M1 vs.
ray-by-ray, e.g. \citealt{Just18}), could cause these discrepancies. However,
we think that both of the results are important in the sense that they
illuminate two faces of the magnetic fields in the explosion dynamics of
non-rotating CCSNe, namely, either in the case of relatively earlier onset of
explosion as pointed out in our study or in the later onset of explosion  as
studied by \citet{Obergaulinger14}.\footnote{In either of the cases, 3D
simulation is apparently needed to draw a robust conclusion to clarify the
roles of magnetic fields on the explosion onset in the non- or slowing-rotating
cores (e.g. \citealt{Muller20b}).}

\section{Magnetic field of proto-neutron star} \label{B-field of PNS}

Shedding light on the dynamics of the MHD core collapse is a main purpose
of this paper. On the other hand, the origin of the magnetic field of the
PNS is also an interesting topic. Especially, the origin of the strong magnetic
field of the magnetar is still under debate  over the three decades (e.g.,
\citealt{Duncan92,Pac92,Thompson93,kouve98}, \citealt{Lai01,Kaspi17} for
a review). Two possible formation scenarios have been proposed to account
for the strong magnetic field of the magnetars. These include a turbulent
dynamo amplification in a rapidly rotating PNS \citep{Thompson93} and a fossil
field hypothesis \citep{Ferrario06, Vink06}. In the latter scenario,
the magnetic field of the progenitor star is amplified mainly due to
the magnetic flux conservation as a result of the  gravitational collapse
of the massive star.

In the present study, we have assumed the non-rotating stellar cores. Based on
the fossil field hypothesis, one can exploratory estimate the strength of
the magnetic field inside the PNS. Since the initial magnetic field is given by
equation (\ref{eq: initial magnetic field}) and uniform ($B=B_0$) inside the
core ($r < 10^{3}$ km), the magnetic field strength of the PNS is estimated
as follows:
\begin{eqnarray}
B_{\rm PNS} \sim 10^{15} {\rm G} \; \Biggl ( \frac{B_0}{10^{12}\ \mathrm{G}} \Biggr )
\Biggl ( \frac{30\ {\rm km}}{r_{\rm PNS}} \Biggr )^2 \;, \label{eq: estimated B_PNS}
\end{eqnarray}
where $r_{\rm PNS}$ denotes the radius of the PNS. From this rough estimate,
one can anticipate that a magnetar-class magnetic field could be formed in
the vicinity of the PNS. Note that the radius of the PNS is defined by the
iso-density surface of $10^{11}$ g/cm$^3$.

\begin{figure}
\begin{center}
\scalebox{0.85}{{\includegraphics{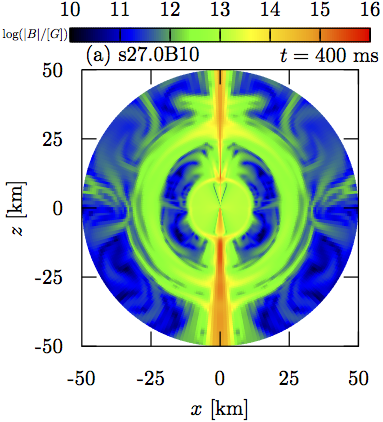}}}
\scalebox{0.85}{{\includegraphics{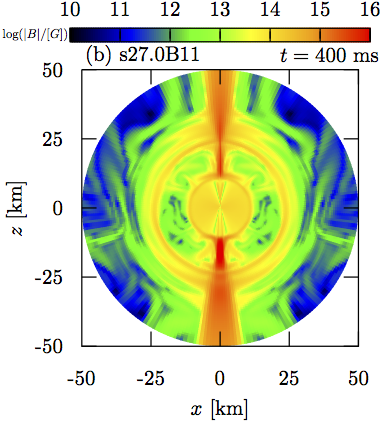}}}
\scalebox{0.85}{{\includegraphics{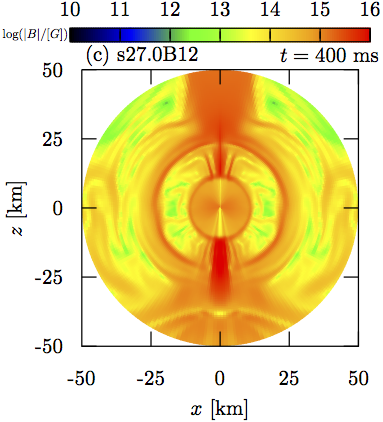}}}
\caption{2D distribution of magnetic field strength for fiducial progenitor
model (s27.0) near at the final simulation time ($t_{\rm pb}  \sim 400$ ms).
Panel (a), (b) and (c) correspond to the cases with $B_0=10^{10}$, $10^{11}$
and $10^{12}$ G, respectively.}
\label{fig8}
\end{center}
\end{figure}

Fig.~\ref{fig8} compares the 2D distribution of the magnetic field strength
for our fiducial runs (s27.0) with $B_0=10^{10}$ G (panel a), $B_0 = 10^{11}$
(panel b) and $B_0 = 10^{12}$ G (panel c), respectively. One can see that the
magnetic field configuration in the central region ($r \lesssim$ 50 km) looks
similar between the three panels. All the three panels reveal a common
shortcoming of our 2D models, namely, the magnetic field is strongest along
the symmetry axis (the $z$-axis) because of the reflective boundary condition
there. In addition, an artificial structure of the magnetic field is also
observed around the symmetry axis in the innermost region ($r<10$ km) where
the calculations are performed in spherical symmetry (see the last paragraph of
Section~\ref{numerical methods}). Having mentioned the shortcoming, let us
focus on model s27.0B12 (panel c).
 
From panel (c), one can see that the field strength inside the region of
$r \lesssim 30$ km is typically $\sim 10^{15}$ G (shown in red), which is
consistent with the rough estimate of equation (\ref{eq: estimated B_PNS})
in the PNS interior. The field strength becomes smaller below the PNS surface
($30 \sim 25$ km $\gtrsim r \gtrsim 12$ km), which is shown in green or
yellowish region in the panel. This region is convectively unstable due to
the negative lepton gradient, whereas the region above is convective stable
due to the positive entropy gradient. In the convective region, it is known
in 2D simulations that the magnetic field is expelled due to the so-called
convective flux expulsion \citep[e.g.][]{Galloway81, Davidson01}. This not only
explains the reduction of the field strength there, but also the accumulation
of the magnetic field above the PNS surface ($r \sim 25 - 30$ km), which can be
seen as the radially ordered field in the range of $25$ km $\lesssim r$
$\lesssim$ 30 km (shown as a red circular band in the panel). This phenomenon
has been already identified in the seminal work by \citet{Obergaulinger14}.
Going more deeper inside ($r \lesssim 12$ km), one can see a rather uniformally
magnetized region ($\sim 10^{15}$ G), which corresponds to the unshocked core,
where the density is close to the nuclear density ($\sim 3 \times 10^{14}$ g/cm$^3$). 

Finally we briefly state some speculations based on our results. The magnetars
are typically observed as anomalous X-ray pulsars or soft gamma repeaters, and
some of them are found in supernova remnants (e.g., \citealt{Kaspi17}). The
X-ray observations of supernova remnants having an association with magnetars
(e.g. Kes 73, N49, and CTB 109) show that the explosion energies of these
objects are comparable to those of canonical CCSNe (e.g., $10^{51}\,{\rm erg}$,
\citealt{Vink06, Nakano17}). That might suggest that these magnetars may not
requre rapid rotators with highly aspherical and energetic jets, but simply
the normal neutrino-driven explosion as the central engine. Our strong initial
field models (B12) would lead to the magnetar-class fields at the surface of
the PNS (Fig. \ref{fig8}), but the explosion occurs by the neutrino heating.
This would be consistent with the finding in the X-ray observations. To prove
this bold speculation, we need to firstly follow a long-term evolution of our
models because the diagnostic explosion energies ($\sim 0.1$ B) still fall
short. Besides, the mass accretion rate of the fallback matter should be
evaluated in the long-term simulation since  the strong flow could bury the
surface magnetic field into the crust \citep{Shigeyama18}. These should be also
verified in the 3D-MHD core-collapse simulations (e.g., \citealt{Winteler12,
Moesta14, Obergaulinger20, Kuroda20}), which could be our next step to be taken. 

\section{Summary and discussion} \label{summary}
We have investigated the impact of the magnetic field on the collapse of
non-rotating stellar cores through 2D axisymmetric MHD simulations. Initially,
$15.0$, $18.4$ or $27.0$ $M_\odot$ presupernova progenitors are threaded by
only the poloidal component of the magnetic field. Since the azimuthal
components of the velocity and magnetic field are zero initially with the 2D
assumption, the evolution of the velocity and magnetic field is restricted
in the poloidal components. We have performed numerical runs for the evolution
of the stellar cores by varying the strength of the magnetic field inside the
core between $B=10^{10}$ and $10^{12}$ G.

The stellar core collapses, and the neutrino-driven explosion occurs in all the
computed models. For the 2D models explored in this work, an intriguing finding
is that the main driver of the explosion is the neutrino heating regardless of
the strength of the initial magnetic field. The magnetic field secondary
contributes to the evolution of the stellar core. The strong magnetic field
prevents the development of the neutrino-driven turbulence in the small scale
compared to the weak magnetic field. This leads to the slow increase of the
turbulent kinetic energy, leading to the slight delay of the neutrino-driven
shock revival.

Finally we shall refer to the limitation of this study. The major limitation of
this work is apparently the 2D assumption, which we shall take mainly for the
sake of our code development which is doable as a first step without using huge
computational resources. However, the explodability of CCSN models has been
shown to be significantly affected by the dimension of the simulation
\citep{Nordhaus10,Hanke12,Dolence13,Takiwaki14,Nakamura19,Nagakura19b,Melson20}.
Especially, 3D purely HD simulations of non-rotating progenitors
indicate negative impacts on the explosion compared to 2D HD simulations.
The magnetic  field may assist the explosion in the 3D simulation of
CCSNe (e.g. \citet{Muller20b}). As already mentioned above, we are
planning to update our code to make 3D-MHD CCSN modeling possible as
recently reported by \cite{Obergaulinger20} and \cite{Muller20b}.
Our method of the ray-by-ray neutrino transport has room to be updated
to the more sophisticated schemes such as the M1-closure scheme
\citep{Shibata11,Oconnor15,Just15,Just18,Kuroda16,Skinner16}, the variable
Eddington factor method \citep{Buras06,Muller10}, and the discrete-angle
($S_n$) method \citep{Liebendoerfer04,Sumiyoshi05,Sumi12,Nagakura14,Nagakura17,
Harada19,Nagakura19a,iwakami20}. Another potential ingredient that can affect
the postbounce dynamics is the treatment of gravity: general relativistic (GR)
simulations (e.g., \citealt{Kuroda12,Kuroda16,Kuroda18,Kuroda20,Roberts16,
Moesta14,Moesta15}) would be particularly important especially for the
collapse of massive stars with high progenitor's compactness (e.g.
\citealt{OO13,Suk16,Ertl16,Ebinger20}). 

We have investigated the dependence of the magnetic field strength on
the dynamics of the core-collapse of the massive star in the context of
the slowly-rotating core. In order to independently clarify the role of
the magnetic field and rotation of the star, we assume no-rotating cores
in this study. The rotational dependence of the core on the dynamics of
the MHD core-collapse is within the scope of our work and will be reported
in our subsequent paper.

Definitively much more elaborate study is needed to unravel the formation
mechanism of magnetars, although we presented a very bold speculation in this
study that our 2D non-rotating and strongly mangetized models could deserve
further investigation toward a future comparison with the observations.
The generation of the magnetar-class field in the context of the PNS dynamo
process \citep{Thompson93} have gained considerable attention recently.
In fact, \cite{Raynaud20} presented the first 3D dynamo simulations, which
showed the generation of the strong magnetic fields in the vicinity of the PNS
(see, also \citealt{Masada20}). The chiral magnetic effects have been reported
to account for the origin of the strong magnetic fields in magnetars
\citep{Yamamoto16,Masada18}. All of these studies require a dedicated 3D MHD
modeling, toward which we have made our first sail in this work.

\section*{Acknowledgements}
We thank K. Nakamura, M. Bugli, Y. Masada, Y. Matsumoto, K. Tomisaka, and
H.-Th. Janka for useful and stimulating discussions.
Numerical computations were
carried out on Cray XC50 at the Center for Computational Astrophysics,
National Astronomical Observatory of Japan and on Cray XC40 at YITP
in Kyoto University. This work was supported by Research Institute of
Stellar Explosive Phenomena at Fukuoka University and the associated project (No. 207002), and also by JSPS KAKENHI
Grant Number (JP17K14260, 
JP17H05206, 
JP17K14306, 
JP17H01130, 
JP17H06364, 
JP18H01212, 
JP18K13591, 
JP19K23443, 
JP20K14473, 
JP20K11851, 
JP20H01941  
and
JP20H00156).  
This research was also supported by MEXT as “Program for Promoting 
researches on the Supercomputer Fugaku” (Toward a unified view of 
he universe: from large scale structures to planets) and JICFuS.

\section*{Data Availability}

The data underlying this article will be shared on reasonable request to the corresponding author.


\bibliographystyle{mnras}
\bibliography{papers} 


\appendix

\section{Full 3D MHD equations in spherical coordinates}\label{equations in spherical coordinate}
The explicit formulae of the ideal MHD equations in spherical
coordinates, neglecting the contribution of the gravity and
the interaction with neutrinos, are
\begin{eqnarray}
\frac{\partial \rho}{\partial t} + \frac{1}{r^2} \frac{\partial (r^2 \rho v_r)}{\partial r}
+\frac{1}{r \sin \theta}\frac{\partial(\rho v_{\theta}\sin\theta)}{\partial \theta}
+\frac{1}{r \sin \theta}\frac{\partial(\rho v_{\phi})}{\partial \phi}
=0\;, \label{eq: mass conservation}
\end{eqnarray}
\begin{eqnarray}
\frac{\partial (\rho v_r)}{\partial t} + \frac{1}{r^2} \frac{\partial [r^2 (\rho v_r v_r + P_t - B_rB_r)]}{\partial r} 
+\frac{1}{r \sin \theta}\frac{\partial[(\rho v_rv_{\theta} - B_rB_{\theta})\sin\theta]}{\partial \theta} 
+\frac{1}{r \sin \theta}\frac{\partial(\rho v_rv_{\phi} - B_rB_{\phi})}{\partial \phi} 
=\frac{\rho(v_{\theta} v_{\theta} + v_{\phi} v_{\phi}) + 2P_t - B_{\theta}B_{\theta} - B_{\phi}B_{\phi}}{r}\;, \label{eq: momentum conservation1}
\end{eqnarray}
\begin{eqnarray}
\frac{\partial (\rho v_\theta)}{\partial t} + \frac{1}{r^2} \frac{\partial [r^2 (\rho v_{\theta} v_r - B_{\theta}B_r)]}{\partial r} 
+\frac{1}{r \sin \theta}\frac{\partial[(\rho v_{\theta}v_{\theta} + P_t - B_{\theta}B_{\theta})\sin\theta]}{\partial \theta}
+\frac{1}{r \sin \theta}\frac{\partial(\rho v_{\theta}v_{\phi} - B_{\theta}B_{\phi})}{\partial \phi}
=\frac{(\rho v_{\phi} v_{\phi} + P_t - B_{\phi}B_{\phi})\cot\theta}{r}
-\frac{\rho v_{\theta} v_r - B_{\theta}B_r}{r}\;, \label{eq: momentum conservation2}
\end{eqnarray}
\begin{eqnarray}
\frac{\partial (\rho v_\phi)}{\partial t} + \frac{1}{r^2} \frac{\partial [r^2 (\rho v_{\phi} v_r - B_{\phi}B_r)}{\partial r} 
+\frac{1}{r \sin \theta}\frac{\partial[(\rho v_{\phi}v_{\theta} - B_{\phi}B_{\theta})\sin\theta]}{\partial \theta} 
+\frac{1}{r \sin \theta}\frac{\partial(\rho v_{\phi}v_{\phi} + P_t - B_{\phi}B_{\phi})}{\partial \phi} 
= -\frac{(\rho v_{\phi} v_{\theta} - B_{\phi}B_{\theta})\cot\theta}{r}
- \frac{\rho v_{\phi} v_r - B_{\phi}B_r}{r}\;, \label{eq: momentum conservation3}
\end{eqnarray}
\begin{eqnarray}
\frac{\partial e}{\partial t} + \frac{1}{r^2} \frac{\partial [r^2 (e + P_t)v_r - r^2 (\mbox{\boldmath $v$} \cdot\mbox{\boldmath $B$}) B_r]]}{\partial r}
+\frac{1}{r \sin \theta}\frac{\partial[(e + P_t)v_{\theta}\sin\theta - (\mbox{\boldmath $v$} \cdot\mbox{\boldmath $B$}) B_{\theta}\sin\theta]}{\partial \theta} 
+\frac{1}{r \sin \theta}\frac{\partial[(e + P_t)v_{\phi} - (\mbox{\boldmath $v$} \cdot\mbox{\boldmath $B$}) B_{\phi}]}{\partial \phi}
=0\;, \label{eq: energy conservation}
\end{eqnarray}
\begin{eqnarray}
\frac{\partial B_r}{\partial t} + \frac{1}{r^2} \frac{\partial (r^2 \psi)}{\partial r}
+\frac{1}{r \sin \theta}\frac{\partial[(B_r v_{\theta} - v_rB_{\theta})\sin\theta]}{\partial \theta} 
+\frac{1}{r \sin \theta}\frac{\partial(B_r v_{\phi} - v_r B_{\phi})}{\partial \phi} 
=\frac{2\psi}{r}\;, \label{eq: induction equation1}
\end{eqnarray}
\begin{eqnarray}
\frac{\partial B_{\theta}}{\partial t} + \frac{1}{r}\frac{\partial[r(B_{\theta} v_r - v_{\theta} B_r)]}{\partial r} 
+\frac{1}{r \sin \theta}\frac{\partial(\psi \sin \theta)}{\partial \theta}
+\frac{1}{r \sin \theta}\frac{\partial(B_{\theta} v_{\phi} - v_{\theta} B_{\phi})}{\partial \phi}
=\frac{\psi \cot\theta}{r}\;, \label{eq: induction equation2}
\end{eqnarray}
\begin{eqnarray}
\frac{\partial B_{\phi}}{\partial t} + \frac{1}{r}\frac{\partial[r(B_{\phi} v_r - v_{\phi} B_r)]}{\partial r}
+\frac{1}{r}\frac{\partial(B_{\phi} v_{\theta} - v_{\phi}B_{\theta})}{\partial \theta}
+\frac{1}{r \sin \theta}\frac{\partial \psi}{\partial \phi}
=0\;, \label{eq: induction equation3}
\end{eqnarray}
\begin{eqnarray}
\frac{\partial \psi}{\partial t} + \frac{1}{r^2} \frac{\partial (r^2 c_h^2 B_r)}{\partial r}
+\frac{1}{r \sin \theta}\frac{\partial[c_h^2 B_{\theta}\sin\theta]}{\partial \theta}
+\frac{1}{r \sin \theta}\frac{\partial (c_h^2 B_{\phi})}{\partial \phi} 
= - \frac{c_h^2}{c_p^2}\psi\;. \label{eq: psi}
\end{eqnarray}
These are the continuity equation (\ref{eq: mass conservation}), the momentum conservation
equations (\ref{eq: momentum conservation1})--(\ref{eq: momentum conservation3}),
the energy conservation equation (\ref{eq: energy conservation}) and the induction
equations (\ref{eq: induction equation1})--(\ref{eq: induction equation3}).
The divergence cleaning method \citep{Dedner02} is implemented in our code to maintain
the numerical errors of solenoidal property of the magnetic field within minimal
levels. The average relative divergence error estimated by
Error$3$($\mbox{\boldmath $B$}$)$^{ave}$ of \citet{Zhang16} is less than 1\% in this work.
Equation (\ref{eq: psi}) and $\psi$ in induction equations
(\ref{eq: induction equation1})--(\ref{eq: induction equation3}) are 
related to the divergence cleaning method. The HLLD scheme \citep{Miyoshi05} is used
to solve equations~(\ref{eq: mass conservation})--(\ref{eq: induction equation3})
in a conservative form. To solve equation \eqref{eq: psi}, the HLLE scheme \citep{einfeldt}
is used.

In addition to the baryon number conservation (equation \ref{eq: mass conservation}), 
we evolve the equation regarding the lepton number conservation (see equation \ref{basic eq6}), 
whose discretization and reconstruction method is the same as that of equation \eqref{eq: mass conservation}.
In the IDSA scheme, the energy equation for the trapped neutrinos needs to be evolved  (see equation \ref{basic eq7}). 
Though the treatment of the advection term of equation \eqref{basic eq7}
is the same as that of equation \eqref{eq: mass conservation}, the left-hand-side of
equation \eqref{basic eq7} includes the term of $p_\nu\nabla \cdot \mbox{\boldmath $v$}$
with $p_\nu = \frac{\rho Z_m}{3}$ the neutrino pressure, which should be treated
as a source term. Then the explicit form of equation \eqref{basic eq6}
in spherical coordinates reads
\begin{eqnarray}
\frac{\partial \rho Z_m }{\partial t} 
+ \frac{1}{r^2} \frac{\partial (r^2 \rho Z_m v_r)}{\partial r}
+\frac{1}{r \sin \theta}\frac{\partial(\rho Z_m v_{\theta}\sin\theta)}{\partial \theta}
+\frac{1}{r \sin \theta}\frac{\partial(\rho Z_m v_{\phi})}{\partial \phi}
=
-\frac{\rho Z_m }{3}\frac{1}{r^2} \frac{\partial (r^2 v_r)}{\partial r}
-\frac{\rho Z_m }{3}\frac{1}{r \sin \theta}\frac{\partial(v_{\theta}\sin\theta)}{\partial \theta}
-\frac{\rho Z_m }{3}\frac{1}{r \sin \theta}\frac{\partial v_{\phi}}{\partial \phi}
\;, \label{eq: neutrino energy conservation}
\end{eqnarray}
noting again that in this Appendix the neutrino-matter interaction term is 
dropped for the sake of brevity.

\begin{table}
\begin{center}
\caption{Summary of the discretization method and the reconstruction of physical variables at the cell surface for conservation equations.}
\begin{tabular}{ccccc}
\hline
discretization method & equation & position of term & discretization formula & reconstruction of physical variables\\
\hline
finite volume 
& (\ref{eq: mass conservation}),
(\ref{eq: momentum conservation1}), (\ref{eq: momentum conservation2}), (\ref{eq: momentum conservation3}) 
(\ref{eq: energy conservation}), (\ref{eq: psi})
& $1$st term
& (\ref{eq: dQdt}) 
&\\
finite volume 
& (\ref{eq: mass conservation}),
(\ref{eq: momentum conservation1}), (\ref{eq: momentum conservation2}), (\ref{eq: momentum conservation3}) 
(\ref{eq: energy conservation}), (\ref{eq: induction equation1}), (\ref{eq: psi})
&$2$nd term & (\ref{eq: FV1}) 
& (\ref{eq: a_L,i}), (\ref{eq: a_R,i}) with (\ref{eq: gamma_i 1}), (\ref{eq: < r >_i 1}) \\ 
finite volume
& (\ref{eq: mass conservation}),
(\ref{eq: momentum conservation1}), (\ref{eq: momentum conservation2}), (\ref{eq: momentum conservation3}) 
(\ref{eq: energy conservation}), (\ref{eq: induction equation2}), (\ref{eq: psi})
&$3$rd term & (\ref{eq: FV2}) 
& (\ref{eq: a_L,j}), (\ref{eq: a_R,j}) \\
finite volume
& (\ref{eq: mass conservation}),
(\ref{eq: momentum conservation1}), (\ref{eq: momentum conservation2}), (\ref{eq: momentum conservation3}) 
(\ref{eq: energy conservation}), (\ref{eq: induction equation3}), (\ref{eq: psi})
&$4$th term & (\ref{eq: FV3}) 
& (\ref{eq: a_L,k}), (\ref{eq: a_R,k}) \\
finite area & (\ref{eq: induction equation1}), (\ref{eq: induction equation2}), (\ref{eq: induction equation3}) 
& $1$st term & (\ref{eq: dBdt}) 
&\\
finite area & (\ref{eq: induction equation1}) & $3$rd term & (\ref{eq: FA1}) & (\ref{eq: a_L,j}), (\ref{eq: a_R,j}) \\
finite area & (\ref{eq: induction equation1}) & $4$th term & (\ref{eq: FA2}) & (\ref{eq: a_L,k}), (\ref{eq: a_R,k})\\
finite area & (\ref{eq: induction equation2}) & $2$nd term & (\ref{eq: FA3}) & (\ref{eq: a_L,i}), (\ref{eq: a_R,i}) with (\ref{eq: gamma_i 2}), (\ref{eq: < r >_i 2}) \\
finite area & (\ref{eq: induction equation2}) & $4$th term & (\ref{eq: FA4}) & (\ref{eq: a_L,k}), (\ref{eq: a_R,k}) \\
finite area & (\ref{eq: induction equation3}) & $2$nd term & (\ref{eq: FA5}) & (\ref{eq: a_L,i}), (\ref{eq: a_R,i}) with (\ref{eq: gamma_i 2}), (\ref{eq: < r >_i 2}) \\
finite area & (\ref{eq: induction equation3}) & $3$rd term & (\ref{eq: FA6}) & (\ref{eq: a_L,k})$^{\rm a}$, (\ref{eq: a_R,k})$^{\rm a}$ \\
\hline
$^{\rm a}$Spacial index $k$ is replaced by $j$.
\label{tableA1}
\end{tabular}
\end{center}
\end{table}

\section{Finite volume and area discretization in spherical coordinates} \label{discretization}

The governing equations for the conservative variables are evolved by the
finite volume and area methods in our code. Both the finite volume and area
methods are used for the induction equations, while only the finite volume
method is used for other equations.

In the spherical coordinates, the conservation equations except the induction
equations are simply described as follows;
\begin{eqnarray}
\frac{\partial Q}{\partial t} + \frac{1}{r^2} \frac{\partial (r^2 F_r)}{\partial r}
+\frac{1}{r \sin \theta}\frac{\partial(F_{\theta}\sin\theta)}{\partial \theta}
+\frac{1}{r \sin \theta}\frac{\partial F_{\phi}}{\partial \phi}
=S\;, \label{eq: conservation equation1}
\end{eqnarray}
where $Q$ is a conservative variable, $F_{r}$, $F_{\theta}$, and $F_{\phi}$ are 
the corresponding flues in the $r$-, $\theta$- and $\phi$- directions and $S$
is a source term. Following the methods proposed in \citet{Li03} and \citet{Mignone14},
we discretize the equation (\ref{eq: conservation equation1}) based on
the finite volume method. Equation (\ref{eq: conservation equation1}) 
is integrated over the cell volume using the Gauss's theorem. The conservative variable
defined at the cell center, $\bar{Q}_{i,j,k}$, is given by the volume average of
$Q$ over the cell volume,
\begin{eqnarray}
\bar{Q}_{i,j,k} = \frac{\int^{i+1/2}_{i-1/2} \int^{j+1/2}_{j-1/2} \int^{k+1/2}_{k-1/2} Q(r, \theta, \phi)\,r^2 \sin \theta\,dr d \theta d \phi}
{\int^{i+1/2}_{i-1/2} \int^{j+1/2}_{j-1/2} \int^{k+1/2}_{k-1/2} r^2 \sin \theta\,d r d \theta d \phi} \;,
\end{eqnarray}
where $i$, $j$ and $k$ stand for the spacial index of the cell center
in the $r$- $\theta$- and $\phi$- directions, respectively. The flux
terms in the left-hand-side of equation \eqref{eq: conservation equation1}
are discretized as follows;
\begin{eqnarray}
\frac{1}{r^2} \frac{\partial (r^2 F_r)}{\partial r} \sim
\frac{{r_{i+1/2}}^2 F_{r, i+1/2, j, k} - {r_{i-1/2}}^2 F_{r, i-1/2, j, k}} {{r_{i+1/2}}^3/3-{r_{i-1/2}}^3/3} \;, \label{eq: FV1}
\end{eqnarray}
\begin{eqnarray}
\frac{1}{r \sin \theta}\frac{\partial(F_{\theta}\sin\theta)}{\partial \theta} \sim
\frac{{r_{i+1/2}}^2/2 - {r_{i-1/2}}^2/2} {{r_{i+1/2}}^3/3-{r_{i-1/2}}^3/3}
\frac{F_{\theta, i, j+1/2, k}\sin \theta_{j+1/2}-F_{\theta, i, j-1/2, k}\sin \theta_{j-1/2}}{\cos \theta_{j-1/2}-\cos \theta_{j+1/2}} \;, \label{eq: FV2}
\end{eqnarray}
\begin{eqnarray}
\frac{1}{r \sin \theta}\frac{\partial F_{\phi}}{\partial \phi} \sim
\frac{{r_{i+1/2}}^2/2 - {r_{i-1/2}}^2/2} {{r_{i+1/2}}^3/3-{r_{i-1/2}}^3/3}
\frac{\theta_{j+1/2} - \theta_{j-1/2}}{\cos \theta_{j-1/2}-\cos \theta_{j+1/2}}
\frac{F_{\phi, i, j, k+1/2} - F_{\phi, i, j, k-1/2}}{\phi_{k+1/2}- \phi_{k+1/2}} \;, \label{eq: FV3}
\end{eqnarray}
where $F_{r, i\pm1/2, j, k}$, $F_{\theta, i, j\pm1/2, k}$
and $F_{\phi, i, j, k\pm1/2}$ are the numerical flues at the cell surface in
each direction (or the interpolated velocity for the source terms in equation
\eqref{eq: neutrino energy conservation}. In this case, the neutrino pressure
is evaluated at the cell center). In addition, $\cot \theta$ and $1/r$ in
the source terms of the discrete conservation equations are replaced like below;
\begin{eqnarray}
\cot{\theta} \sim
\frac{\sin \theta_{j+1/2} - \sin \theta_{j-1/2}}{\cos \theta_{j-1/2}-\cos \theta_{j+1/2}} \;,
\end{eqnarray}
\begin{eqnarray}
\frac{1}{r} \sim
\frac{{r_{i+1/2}}^2/2 - {r_{i-1/2}}^2/2} {{r_{i+1/2}}^3/3-{r_{i-1/2}}^3/3} \;.
\end{eqnarray}
Physical variables in the source terms are also evaluated by the cell-volume-averaged
values. The time derivative of equation (\ref{eq: conservation equation1}) is
discretized forward in time:
\begin{eqnarray}
\frac{\partial Q}{\partial t} \sim
\frac{\bar{Q}^{n+1}_{i,j,k} - \bar{Q}^{n}_{i,j,k}}{\Delta t} \;, \label{eq: dQdt}
\end{eqnarray}
where the superscript $n$ stands for the number of time steps and $\Delta t$
is a step size. This formulation leads to the first-order accuracy in time,
which will be extended in higher-order accuracy (Matsumoto et al. in 
preparation).

In the induction equations (\ref{eq: induction equation1})--(\ref{eq: induction equation3}),
the terms including $\psi$ are also discretized based on the finite volume
method. However, considering the magnetic flux conservation, other terms should
be better discretized based on the finite area method, namely, by integrating
these terms over the cell area using the Stokes' theorem. The terms discretized
by the finite area method are listed in Table \ref{tableA1}. The first terms
in the left-hand-side of the induction equations are time derivatives of each
component of the magnetic field. Here, we introduce the area average of the variables
over the cell areas defined at the cell center, 
\begin{eqnarray}
\bar{B}_{r, i, j, k} = \frac{\int^{j+1/2}_{j-1/2} \int^{k+1/2}_{k-1/2} B_{r}(\theta, \phi)\, r^2 \sin \theta d \theta d \phi}
{\int^{j+1/2}_{j-1/2} \int^{k+1/2}_{k-1/2} r^2 \sin \theta d \theta d \phi} \;, 
\end{eqnarray}
\begin{eqnarray}
\bar{B}_{\theta, i, j, k} = \frac{\int^{i+1/2}_{i-1/2} \int^{k+1/2}_{k-1/2} B_{\theta}(r, \phi)\, r \sin \theta dr d \phi}
{\int^{i+1/2}_{i-1/2} \int^{k+1/2}_{k-1/2} r \sin \theta dr d \phi} \;, 
\end{eqnarray}
\begin{eqnarray}
\bar{B}_{\phi, i, j, k} = \frac{\int^{i+1/2}_{i-1/2} \int^{j+1/2}_{j-1/2} B_{\phi}(r, \theta)\, r dr d \theta}
{\int^{i+1/2}_{i-1/2} \int^{j+1/2}_{j-1/2} r dr d \theta} \;, 
\end{eqnarray}
respectively. The flux in the conservation form of the induction equation is
related to the electric field defined as:
\begin{eqnarray}
E_{r} = \frac{B_{\theta}v_{\phi} - v_{\theta}B_{\phi}}{c} \; ,
\end{eqnarray}
\begin{eqnarray}
E_{\theta} = \frac{B_{\phi}v_{r} - v_{\phi}B_{r}}{c} \; ,
\end{eqnarray}
\begin{eqnarray}
E_{\phi} = \frac{B_{r}v_{\theta} - v_{r}B_{\theta}}{c} \; ,
\end{eqnarray}
where $c$ is the speed of light. The flux terms of the induction equations
are then discretized as follows;
\begin{eqnarray}
\frac{1}{r \sin \theta}\frac{\partial(cE_{\phi}\sin\theta)}{\partial \theta} \sim
\frac{r_{i+1/2} - r_{i-1/2}}{r^2_{i+1/2}/2 - r^2_{i-1/2}/2}
\frac{cE_{\phi, i, j+1/2, k} \sin\theta_{j+1/2} - cE_{\phi, i, j-1/2, k} \sin\theta_{j-1/2}}{\cos \theta_{j-1/2} - \cos \theta_{j+1/2}} \;, \label{eq: FA1}
\end{eqnarray}
\begin{eqnarray}
\frac{1}{r \sin \theta}\frac{\partial (-cE_{\theta})}{\partial \phi} \sim
\frac{r_{i+1/2} - r_{i-1/2}}{r^2_{i+1/2}/2 - r^2_{i-1/2}/2}
\frac{\theta_{j+1/2} - \theta_{j-1/2}}{\cos \theta_{j-1/2} - \cos \theta_{j+1/2}}
\frac{-cE_{\theta, i, j, k+1/2} + cE_{\theta, i, j, k-1/2}}{\phi_{k+1/2} - \phi_{k-1/2}} \;, \label{eq: FA2}
\end{eqnarray}
\begin{eqnarray}
\frac{1}{r} \frac{\partial (-r cE_{\phi})}{\partial r} \sim
\frac{- r_{i+1/2} cE_{\phi, i+1/2, j, k} + r_{i-1/2} cE_{\phi, i-1/2, j, k}}{r^2_{i+1/2}/2 - r^2_{i-1/2}/2} \; , \label{eq: FA3}
\end{eqnarray}
\begin{eqnarray}
\frac{1}{r \sin \theta} \frac{\partial cE_{r}}{\partial \phi} \sim
\frac{r_{i+1/2} - r_{i-1/2}}{r^2_{i+1/2}/2 - r^2_{i-1/2}/2}
\frac{\theta_{j+1/2} - \theta_{j-1/2}}{\cos \theta_{j-1/2} - \cos \theta_{j+1/2}}
\frac{cE_{r, i, j, k+1/2} -c E_{r, i, j, k-1/2}}{\phi_{k+1/2} - \phi_{k-1/2}} \;, \label{eq: FA4}
\end{eqnarray}
\begin{eqnarray}
\frac{1}{r} \frac{\partial (r cE_{\theta})}{\partial r} \sim
\frac{r_{i+1/2} cE_{\theta, i+1/2, j, k} - r_{i-1/2} cE_{\theta, i-1/2, j, k}}{r^2_{i+1/2}/2 - r^2_{i-1/2}/2} \; , \label{eq: FA5}
\end{eqnarray}
\begin{eqnarray}
\frac{1}{r} \frac{\partial (-cE_{r})}{\partial \theta} \sim
\frac{r_{i+1/2} - r_{i-1/2}}{r^2_{i+1/2}/2 - r^2_{i-1/2}/2}
\frac{-cE_{r, i, j+1/2, k} + cE_{r, i, j-1/2, k}}{\theta_{j+1/2} - \theta_{j+1/2}} \; . \label{eq: FA6}
\end{eqnarray}
The time derivative of the induction equation is also discretized forward
in time:
\begin{eqnarray}
\frac{\partial B_{s}}{\partial t} \sim
\frac{\bar{B}^{n+1}_{s,i,j,k} - \bar{B}^{n}_{s,i,j,k}}{\Delta t} \;, \label{eq: dBdt}
\end{eqnarray}
where $s$ represents the direction ($r, \theta, \phi$). 

The numerical fluxes at the cell surface are normally computed by solving
a Riemann problem between the discontinuous states across the cell interfaces.
In our code, an approximate Riemann solver  \citep[HLLD;][]{Miyoshi05}
is used to estimate the numerical fluxes.

On the choice of the parameters in equation \eqref{eq: psi}, $c_h$
represents the propagation speed of the numerical error of the deviation
from $\nabla \cdot\mbox{{\boldmath $B$}}=0$. For the value, the largest
eigenvalue in the global computational domain is often taken (e.g., \citealt{Zhang16}):
\begin{eqnarray}
c_h &= \max_{i,j,k}\left(c_s+|v_r|,c_s+|v_\theta|,c_s+|v_\phi| \right) \;,
\end{eqnarray}
where $c_s$ is the total sound speed that includes the effect of the magnetic
fields. Note that we can also evaluate it from the Courant--Friedrichs--Lewy (CFL)
condition \citep{Matsumoto19}. In order to determine $c_p$, we have to take
care the effect of non-uniform grid size that employed in the simulations.
Following \cite{Zhang16}, we set parameter $\alpha = \Delta h_{i,j,k} c_h/c_p^2=0.1$
and $c_p$ is chosen to satisfy this relation. Here $\Delta h_{i,j,k}$ is
the minimum grid size that is taken locally:
\begin{eqnarray}
\Delta h_{i,j,k} &= \min \left(\Delta r_i, r_i \Delta \theta_j, r_i\sin\theta_j \Delta \phi_k\right) \;,
\end{eqnarray}
where $\Delta r_i = r_{i+1/2} - r_{i-1/2}$,
$\Delta \theta_j = \theta_{j+1/2} - \theta_{j-1/2}$,
$\Delta \phi_k = \phi_{k+1/2} - \phi_{k-1/2}$.
Practically we need only the damping rate (the coefficient of $\psi$
in RHS of equation \ref{eq: psi}), 
namely, $c_h^2/c_p^2= \alpha c_h / \Delta h_{i,j,k}$, which indeed 
has a unit of $1/\rm {s}$, and do not
need to evaluate $c_p$ itself.

\section{Reconstruction of volume and area averages} \label{reconstruction}
In order to achieve the high-order accuracy in space, the reconstruction
of physical variables at the cell surface is necessary in the procedure
of the estimation for the numerical fluxes. The piecewise linear method   
(PLM) of reconstruction (second-order) is implemented in our code.
Note that since both finite volume and area methods are used in our code,
the reconstructions of volume and area averages are necessary. 
Following the method of \citet{Skinner10} and \citet{Mignone14},
the reconstruction of the volume average is obtained. First, we
consider the reconstruction in the $r$-direction. Let $a_i$ be a
cell-volume-averaged physical value inside zone $i$ defined at
the cell center, that is,
\begin{eqnarray}
a_i = \frac{\int^{r_{i+1/2}}_{r_{i-1/2}} a(r)\, r^2 d r}{\int^{r_{i+1/2}}_{r_{i-1/2}} r^2 d r} \;.
\end{eqnarray}
The physical variables at the left and right interfaces inside the zone $i$
are then given by 
\begin{eqnarray}
a_{L, i} = a_i - \frac{1}{2} \Delta a_i (1+\gamma_i) \;, \label{eq: a_L,i}
\end{eqnarray}
\begin{eqnarray}
a_{R, i} = a_i + \frac{1}{2} \Delta a_i (1-\gamma_i) \;, \label{eq: a_R,i}
\end{eqnarray}
where 
\begin{eqnarray}
\gamma_i = \frac{\Delta r_i}{r_i} \frac{4}{12 + \Delta r^2_i / r^2_i} \label{eq: gamma_i 1}
\end{eqnarray}
is a correction factor for curvature in the spherical coordinates and
\begin{eqnarray}
\Delta r_i = r_{i+1/2} - r_{i-1/2} \;.
\end{eqnarray}
Here, $\Delta a_i$ is the difference slope of the physical variable.
In our code, the modified van Leer (VL) limiter proposed in \citet{Mignone14}
is implemented for the slope limiter to achieve the total variation diminishing (TVD);
\begin{eqnarray}
\Delta a_i = \Delta a_i^F \varphi^{\rm VL} (\upsilon) \; , \label{eq: difference slope}
\end{eqnarray}
where
\begin{eqnarray}
\upsilon = \frac{\Delta a_i^B}{\Delta a_i^F} \; .
\end{eqnarray}
The forward- and backward-difference slopes are given by
\begin{eqnarray}
\Delta a_i^F = (a_{i+1} - a_{i}) \frac{\Delta r_i}{\langle r \rangle_{i+1} - \langle r \rangle_{i}} \;,
\end{eqnarray}
\begin{eqnarray}
\Delta a_i^B = (a_{i} - a_{i-1}) \frac{\Delta r_i}{\langle r \rangle_{i} - \langle r \rangle_{i-1}} \;,
\end{eqnarray}
where $\langle r \rangle_i$ is the cell-volume-averaged value inside zone $i$;
\begin{eqnarray}
\langle r \rangle_i = \frac{\int^{r_{i+1/2}}_{r_{i-1/2}} r^3 d r}{\int^{r_{i+1/2}}_{r_{i-1/2}} r^2 d r}
= r_i + \frac{2 r_i \Delta r^2}{12r^2_i + \Delta r^2} \;. \label{eq: < r >_i 1}
\end{eqnarray}
The modified VL limiter is defined as follows \citep[see][for details]{Mignone14};
\begin{eqnarray}
\varphi^{\rm VL} (\upsilon)= \left\{ 
    \begin{array}{cl}
         \frac{\upsilon(c^F_i \upsilon + c^B_i)}{\upsilon^2 + (c^F_i + c^B_i - 2)\upsilon +1} & {\rm for} \; \upsilon \ge 0 \;, \\
         0 &  {\rm for} \; \upsilon = 0 \;,
    \end{array} \right. \label{eq: van Leer limiter}
\end{eqnarray}
where
\begin{eqnarray}
c^F_i = \frac{\langle r \rangle_{i+1} - \langle r \rangle_{i}}{r_{i+1/2} - \langle r \rangle_i} \;, 
\end{eqnarray}
\begin{eqnarray}
c^B_i = \frac{\langle r \rangle_{i} - \langle r \rangle_{i-1}}{\langle r \rangle_i - r_{i-1/2}} \; .
\end{eqnarray}
We employ the modified VL limiter in all  the directions for the reconstructions
of both volume and area averages.

In the $\theta$-direction, the cell-volume-averaged physical value
inside zone $j$ defined at the cell center is given by
\begin{eqnarray}
a_j = \frac{\int^{\theta_{j+1/2}}_{\theta_{j-1/2}} a(\theta) \sin \theta d \theta}{\int^{\theta_{j+1/2}}_{\theta_{j-1/2}} \sin \theta d \theta} \;. \label{eq: a_j}
\end{eqnarray}
The physical variables at the left and right interfaces inside the zone $j$
are then given by 
\begin{eqnarray}
a_{L, j} = a_j - \frac{1}{2} \Delta a_j (1+\gamma_j) \;, \label{eq: a_L,j}
\end{eqnarray}
\begin{eqnarray}
a_{R, j} = a_j + \frac{1}{2} \Delta a_j (1-\gamma_j) \;, \label{eq: a_R,j}
\end{eqnarray}
where
\begin{eqnarray}
\gamma_j = \frac{\cos\theta_j}{\sin\theta_j}
\biggl [
\frac{2}{\Delta \theta_j}
-
\frac{\cos (\Delta \theta_j/2)}
     {\sin (\Delta \theta_j/2)}
\biggr ] \; \label{eq: gamma_j}
\end{eqnarray}
is also the correction factor for curvature in the spherical coordinates and
\begin{eqnarray}
\Delta \theta_j = \theta_{j+1/2} - \theta_{j-1/2} \;.
\end{eqnarray}
Here, $\Delta a_j = \Delta a_j^F \varphi^{\rm VL}$ is the difference
slope of the physical variable in the $\theta$-direction. The forward- and
backward-difference slopes used in the VL limiter are given by
\begin{eqnarray}
\Delta a_j^F = (a_{j+1} - a_{j}) \frac{\Delta \theta_j}{\langle \theta \rangle_{j+1} - \langle \theta \rangle_{j}} \; ,
\end{eqnarray}
\begin{eqnarray}
\Delta a_j^B = (a_{j} - a_{j-1}) \frac{\Delta \theta_j}{\langle \theta \rangle_{j} - \langle \theta \rangle_{j-1}} \;, 
\end{eqnarray}
where $\langle \theta \rangle_j$ is the cell-volume-averaged value inside zone $j$;
\begin{eqnarray}
\langle \theta \rangle_j = \frac{\int_{\theta_{j-1/2}}^{\theta_{j+1/2}} \theta \sin\theta d \theta}
{\int_{\theta_{j-1/2}}^{\theta_{j+1/2}} \sin\theta d \theta}
= \theta_j+\frac{\cos\theta_j}{\sin\theta_j}\biggl [
1-\frac{\Delta\theta_j}{2}
\frac{\cos (\Delta \theta_j/2)}
     {\sin (\Delta \theta_j/2)}
\biggr ] \; .
\end{eqnarray}
The coefficients $c^F_i$ and $c^B_i$ in equation (\ref{eq: van Leer limiter})
are replaced by
\begin{eqnarray}
c^F_j = \frac{\langle \theta \rangle_{j+1} - \langle \theta \rangle_{j}}{\theta_{j+1/2} - \langle \theta \rangle_j} \;
\end{eqnarray}
and
\begin{eqnarray}
c^B_j = \frac{\langle \theta \rangle_{j} - \langle \theta \rangle_{j-1}}{\langle \theta \rangle_j - \theta_{j-1/2}} \; ,
\end{eqnarray}
respectively.

In the $\phi$-direction, the reconstruction of the physical variable
is given by the same formula as the Cartesian coordinates. The
cell-volume-averaged physical value inside zone $z$ defined at
the cell center is given by
\begin{eqnarray}
a_k = \frac{\int^{\phi_{k+1/2}}_{\phi_{k-1/2}} a(\phi) \, d \phi}{\int^{\phi_{k+1/2}}_{\phi_{k-1/2}} d \phi} \;.
\end{eqnarray}
The physical variables at the left and right interfaces inside the zone
$k$ are reconstructed as follows;
\begin{eqnarray}
a_{L, k} = a_k - \frac{1}{2} \Delta a_k \;, \label{eq: a_L,k}
\end{eqnarray}
\begin{eqnarray}
a_{R, k} = a_k + \frac{1}{2} \Delta a_k \;, \label{eq: a_R,k} 
\end{eqnarray}
where $\Delta a_k = \Delta a_k^F \varphi^{\rm VL}$ is the difference
slope of the physical variable in the $\phi$-direction. These are simply
linear interpolations without correction factors for curvature. The forward-
and backward-difference slopes used in the VL limiter are defined by
\begin{eqnarray}
\Delta a_k^F =  a_{k+1} - a_{k} \; ,
\end{eqnarray}
\begin{eqnarray}
\Delta a_k^B = a_{k} - a_{k-1} \;.
\end{eqnarray}
Note that the coefficients $c^F_k$ and $c^B_k$ of the VL limiter are $2$
in this case.

Next, we consider the reconstruction of area averaged values. 
The cell-area-averaged physical value inside zone $i$ along the 
$r$-direction defined at the cell center is given by
\begin{eqnarray}
a_i = \frac{\int^{r_{i+1/2}}_{r_{i-1/2}} a(r) \, r d r}{\int^{r_{i+1/2}}_{r_{i-1/2}} r d r} \;.
\end{eqnarray}
This is the same formula for the case with the reconstruction of
volume averaged value along the radial direction in the cylindrical coordinates
\citep[see][]{Skinner10}. The physical variables at the left and right interfaces
inside the zone $i$ are given by equations (\ref{eq: a_L,i}) and
(\ref{eq: a_R,i}), respectively. However, the correction factor
for curvature is
\begin{eqnarray}
\gamma_i = \frac{\Delta r_i}{6r_i} \; \label{eq: gamma_i 2}
\end{eqnarray}
in this case. In the reconstruction of area averaged values, we also employ
the VL limiter (\ref{eq: difference slope}) and (\ref{eq: van Leer limiter}).
The cell-area-averaged value of $r$ inside zone $i$ used in the VL
limiter is given by
\begin{eqnarray}
\langle r \rangle_i = r_i + \frac{(\Delta r_i)^2}{12r_i} \; . \label{eq: < r >_i 2}
\end{eqnarray}

In the $\theta$-direction, there are two types of reconstructions.
It depends on the direction of the area. When the area is perpendicular
to the  $r$-direction, the cell-area-averaged physical value inside zone
$j$ defined at the cell center is given by
\begin{eqnarray}
a_j = \frac{\int^{\theta_{j+1/2}}_{\theta_{j-1/2}} a(\theta) \sin \theta d \theta}{\int^{\theta_{j+1/2}}_{\theta_{j-1/2}} \sin \theta d \theta} \;.
\end{eqnarray}
This is the same formula as the cell-volume-averaged physical value
inside zone $j$ (\ref{eq: a_j}). Therefore, the reconstructions of the
physical value at the cell surfaces are given by equations (\ref{eq: a_L,j})
and (\ref{eq: a_R,j}). On the other hand, when the area is perpendicular
to the $\phi$-direction, 
\begin{eqnarray}
a_j = \frac{\int^{\theta_{j+1/2}}_{\theta_{j-1/2}} a(\theta) \, d \theta}{\int^{\theta_{j+1/2}}_{\theta_{j-1/2}} d \theta} \;.
\end{eqnarray}
This simply results in the linear interpolation at the cell surface like
equation (\ref{eq: a_L,k}) or (\ref{eq: a_R,k}) although the spacial
index $k$ in the equations is replaced by $j$.

In the $\phi$-direction, the reconstruction of the physical variable is the same
as the case with the volume-averaged reconstruction. It is simply the linear
interpolation of the physical variable and given by equation
(\ref{eq: a_L,k}) or (\ref{eq: a_R,k}).

For a concise summary, the physical variables reconstructed by the volume-
and area-averaged methods are listed in Table \ref{tableA1}.

\section{Blast wave in a strongly magnetized medium}\label{code test1}
\begin{figure}
\begin{center}
\includegraphics[width=.4\linewidth]{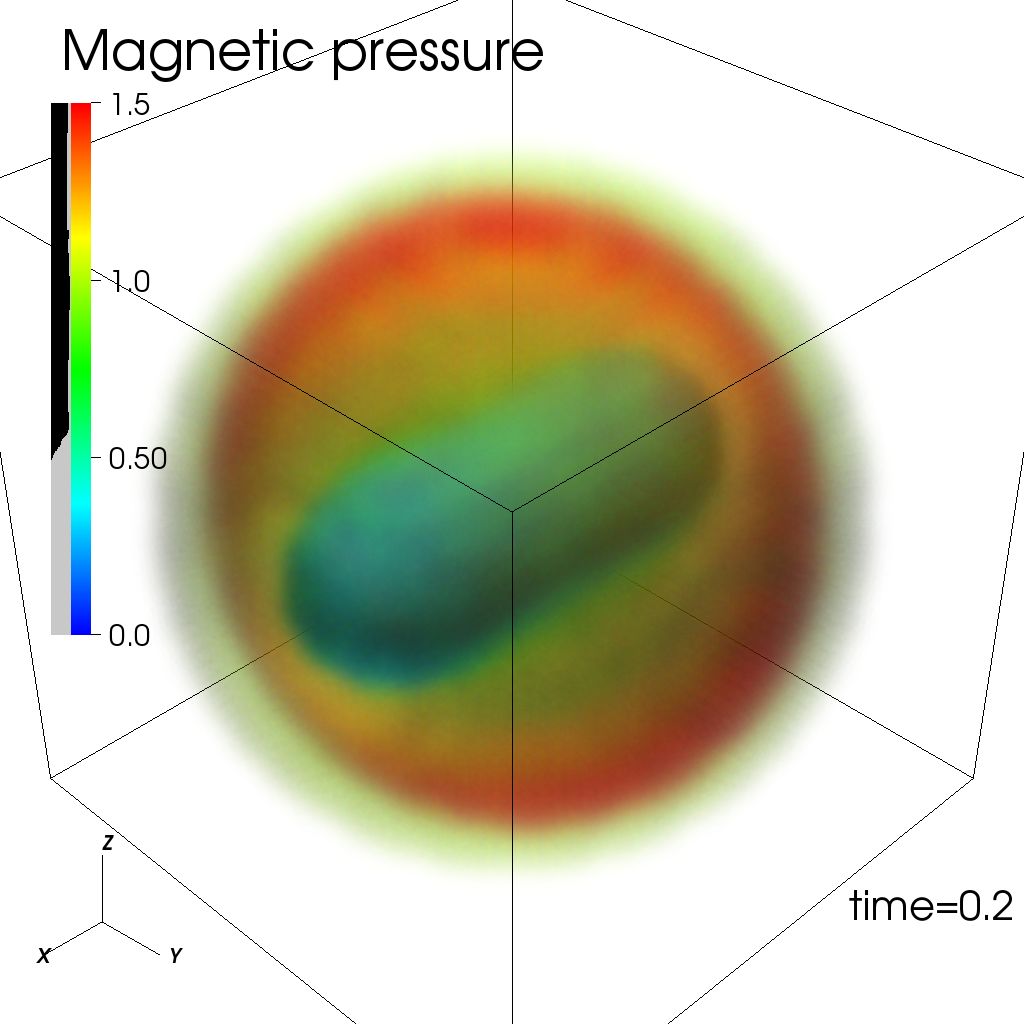}
\includegraphics[width=.4\linewidth]{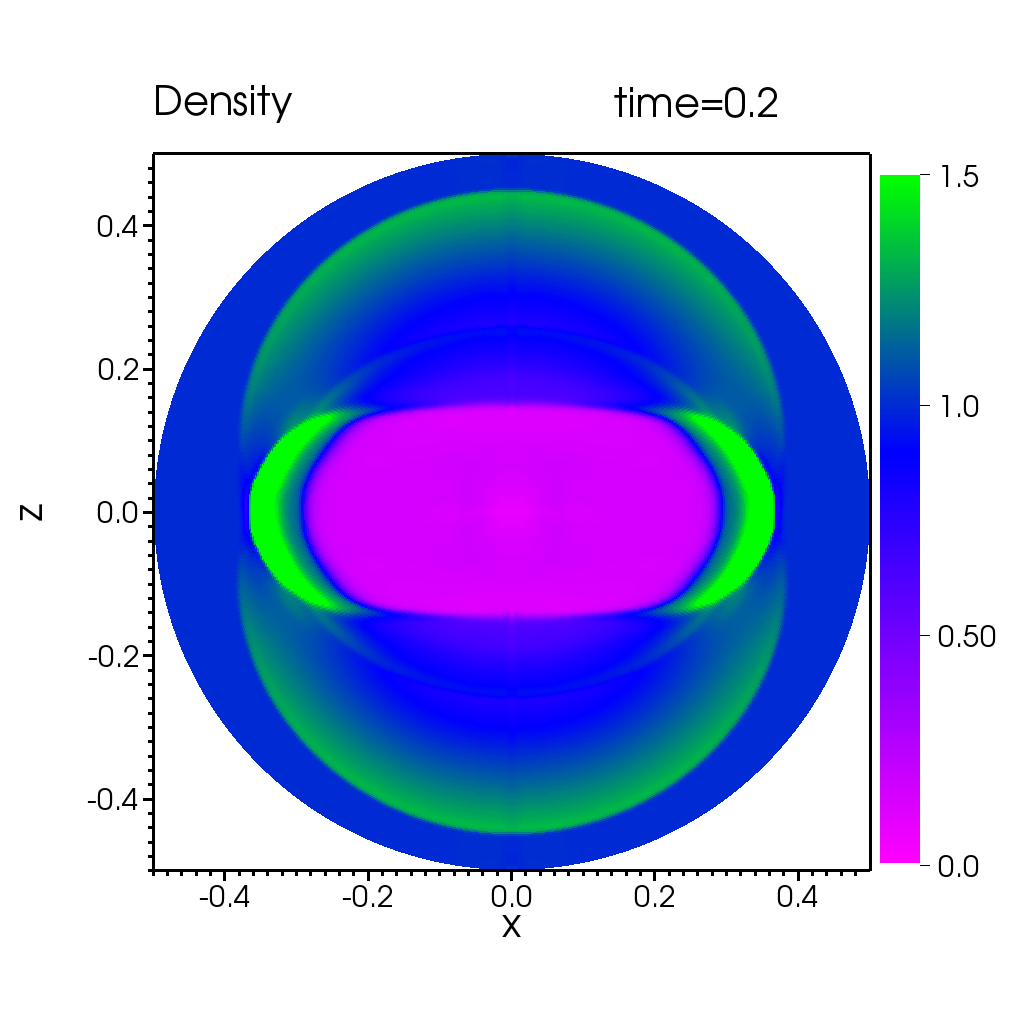}\\
\includegraphics[width=.4\linewidth]{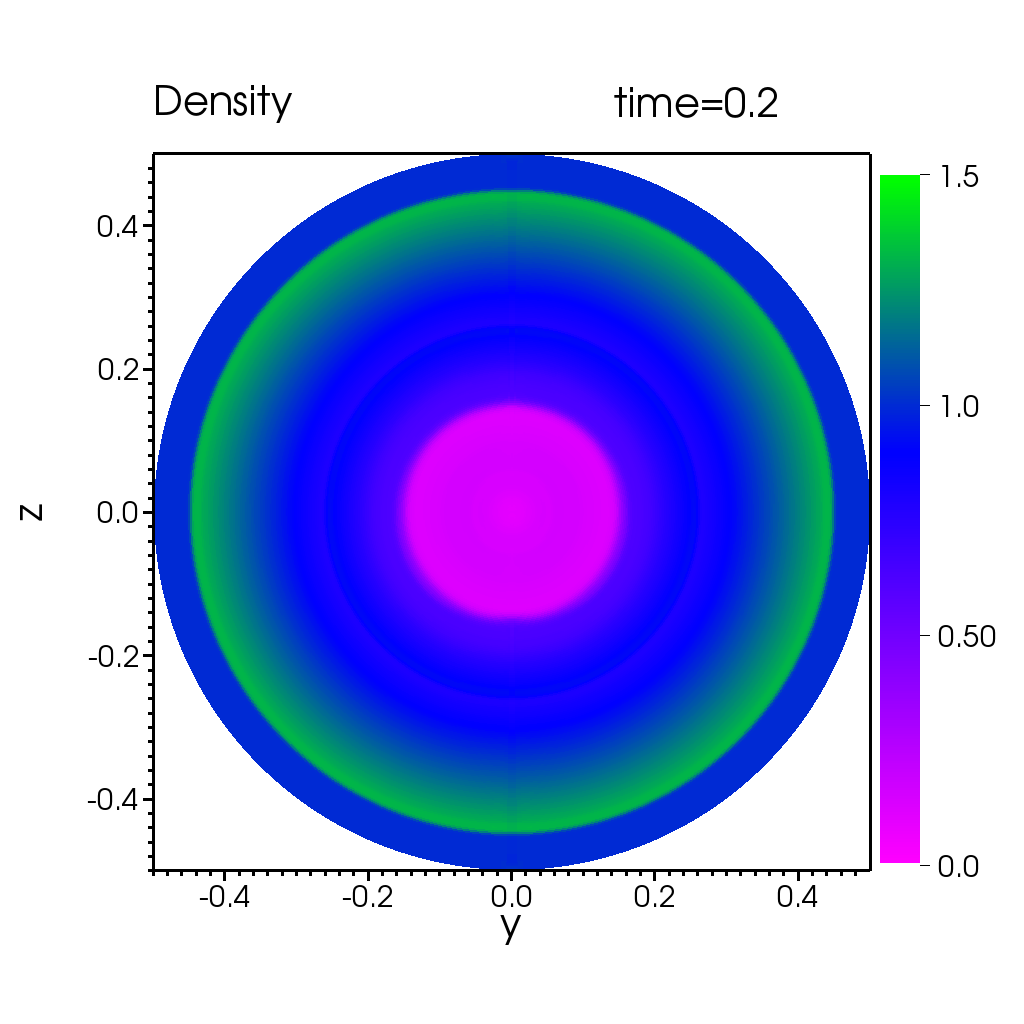}
\includegraphics[width=.4\linewidth]{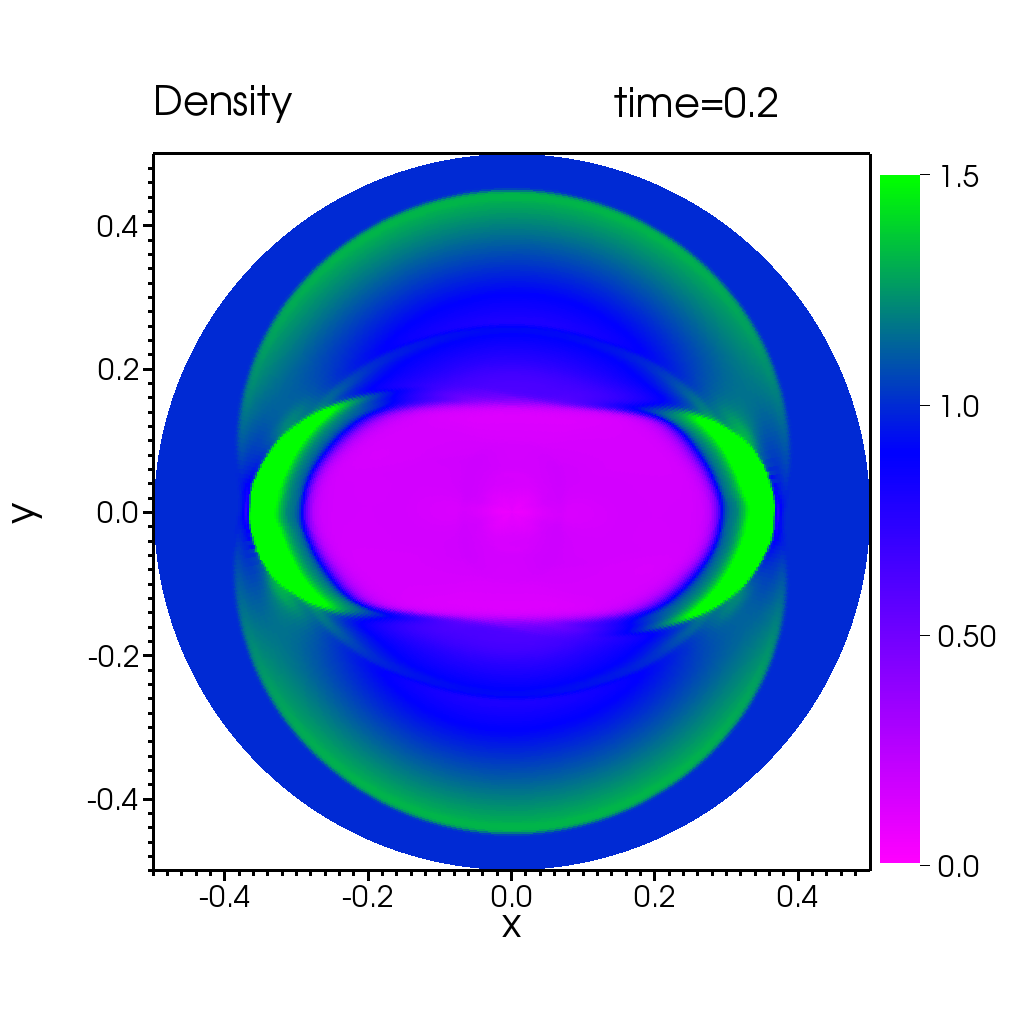}
\caption{Snapshot of the 3D blast wave. Top left: The volume-rendered image of
the magnetic pressure. Top right, bottom left and bottom right panels show the
density structure in 2D slice of $y=0$, $x=0$ and $z=0$, respectively. The
uniform magnetic field is initially imposed in $x$-direction. }
\label{fig:mhdblast}
\end{center}
\end{figure}
For a code verification, we demonstrate a test of an MHD blast wave problem
in 3D. This test is famous and performed in many previous works with similar
setups \citep{Londrillo00, Stone08, Stone09, Skinner10}. The initial condition
that we adopt is as follows. We set a static background of $\rho=1.0, p=0.1$.
The background is magnetized as $B_x=1.0$ (in our definition, $B$ is already
normalized by $\sqrt{4\pi}$). In the background, we put a hot gas of $p=10$
in the region of $r<0.1$. The polytropic index of EOS, $\Gamma$, is set to
$\frac{5}{3}$. The domain of $[0,0.5]\times[0,\pi]\times[0,2\pi]$ of spherical
polar coordinate is uniformly covered by $200\times200\times400$ grids,
respectively.

The time snapshot of $t=0.2$ is shown in Fig. \ref{fig:mhdblast}.
The top left panel is the volume-rendered image of the magnetic pressure.
The blast wave spherically propagates outward. The shock front is located
with red sphere and green envelop in the figure. The central region has
higher pressure and lower magnetic pressure. The region is significantly
collimated due to the effect of the magnetic field. It is also important
to check the density distribution. The top right, bottom left and bottom right
panels are the 2D slice of $y=0$, $x=0$ and $z=0$ planes, respectively.
The structure of the density is quite similar to that in the other works
(see Fig. 36 of \citealt{Stone08}). The shock front corresponds to the
high-density outer ring (shown as green) and that is almost spherical.
The central low density region is the consequence of the rarefaction
which propagates inward. The crescent density structure at the front of
the shock in the $x$-direction is typically  seen in this test.

The evolution of the shock does not depend on the coordinate system.
Though the structure of the mesh is quite different in the $x$-$z$ ($r$-$\theta$)
plane in the top right panel and  the $x$-$y$ ($r$-$\phi$) plane in the bottom
right panel, the density structures of them are quite similar. Such a coordinate
independent feature would support the correctness of our code implementation.

\section{Resolution study of the shock revival with magnetic fields} \label{resolution study}
\begin{figure}
\begin{center}
\scalebox{0.9}{{\includegraphics{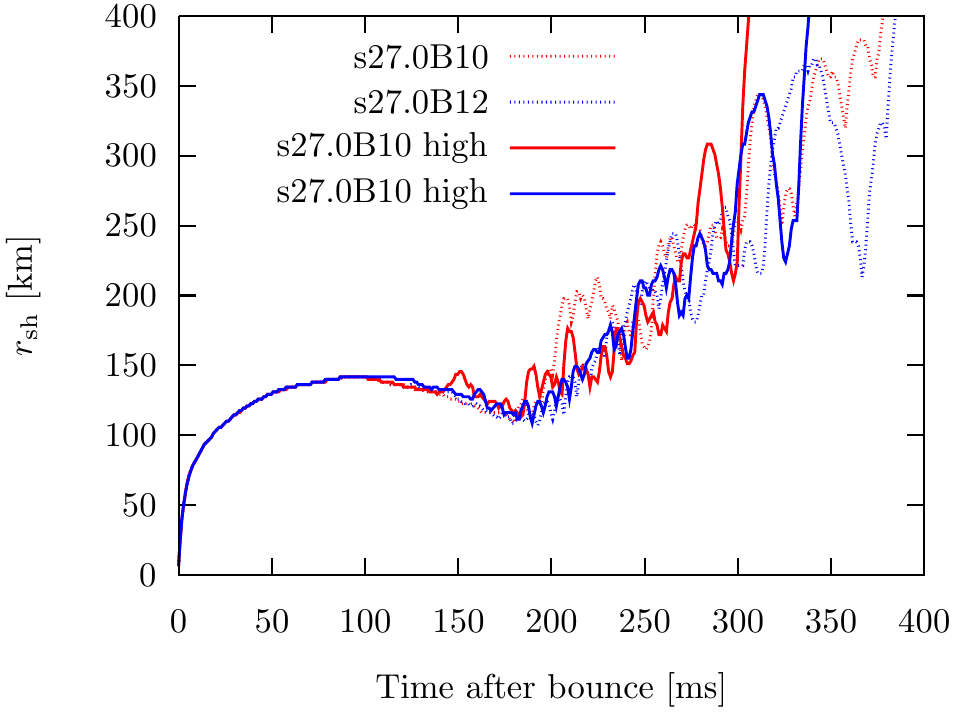}}}
\caption{Temporal evolution of the shock radii for our fiducial progenitor model
(s27.0) with high numerical resolution (see the text). Solid, red and blue lines represent the weak
($B_0=10^{10}$ G) and strong ($B_0=10^{12}$ G) magnetic field model,
respectively. For the reference, the results of fiducial runs are 
demonstrated by doted lines.}
\label{figE1}
\end{center}
\end{figure}
The influence of the numerical resolution on the evolution of the MHD
core collapse is investigated by changing the grid spacing in the
$\theta$-direction. The number of grid points in the $\theta$-direction
for fiducial runs is $128$ as described in Section \ref{numerical methods}.
We run s27.0B10 and s27.0B12 models with the resolution of
$\Delta \theta = \pi / 256$. The number of grid points in
high resolution calculations is twice as larger as the fiducial runs.

Fig.~\ref{figE1} shows the temporal evolution of the shock radius
in the high resolution runs. Solid, red and blue lines correspond to
the models s27.0B10 and s27.0B12, respectively. For the reference,
the results of fiducial runs for both models are shown by thin dotted lines.

In both weak (s27.0B10) and strong (s27.0B12) magnetic field models,
the shock radius in the high resolution run expands fast compared to
that in the fiducial run. Solid lines reach 400 km faster than doted
lines as shown in Fig~\ref{figE1}. This result is consistent with
a recent resolution study for the supernova simulation \citep{Melson20}.
The higher angular resolution provides more favorable explosion
conditions.

Also in the high resolution runs, the  delay of the onset of the
shock re-expansion with the stronger magnetic field models is observed at
$t_{\rm pb} \sim 200$ ms. The onset of the shock expansion in the weak
magnetic field model is faster than that in the strong field model.
As mentioned in Section~\ref{results}, this tendency has been also
observed in the fiducial resolution runs. Our resolution study demonstrates
that the delay of the shock revival occurs regardless of the angular
resolution in the $\theta$-direction although more careful studies
covering the wide range of resolution are necessary in order to draw
a robust conclusion.
\label{lastpage}
\end{document}